\documentclass[11pt, a4paper, logo, copyright, nonumbering]{xiaomi}

\usepackage[authoryear, sort&compress, round]{natbib}
\usepackage{dblfloatfix}
\usepackage{ulem}
\usepackage{caption}
\usepackage{dramatist}
\usepackage{xspace}
\usepackage{pifont}
\usepackage{multirow}
\usepackage{tcolorbox}
\usepackage{xltabular}
\usepackage{longtable}
\usepackage{hyperref}
\usepackage{wrapfig}
\usepackage{graphicx}

\usepackage{amsfonts}
\usepackage{amsmath}
\usepackage{amssymb}
\usepackage{lineno}
\usepackage{multirow}
\usepackage{adjustbox}
\usepackage{float}
\usepackage[bottom]{footmisc}

\usepackage{CJKutf8}
\usepackage{subcaption}
\usepackage{setspace}
\usepackage{makecell}
\usepackage{graphicx}
\usepackage{multicol}
\usepackage{algorithm}
\usepackage[noend]{algpseudocode}

\defcitealias{nondeter}{T. M. Lab}

\newcommand{\sysname}{ARL-Tangram\xspace}

\newcommand{\parabf}[1]{\smallskip\noindent\textbf{#1}}

\newcommand{\paraf}[1]{\noindent\textbf{#1}}
\newcommand{\cut}[1]{}

\definecolor{exitcolor}{RGB}{200,60,60}      % red for exits
\definecolor{commentcolor}{RGB}{140,140,140} % light gray for comments
\newcommand{\algcomment}[1]{\hfill{\scriptsize\textcolor{commentcolor}{// #1}}}
\definecolor{xiaomiorange}{HTML}{FF6901}

\begin{abstract}

Agentic reinforcement learning (RL) has emerged as a transformative workload in cloud clusters, enabling large language models (LLMs) to solve complex problems through interactions with real world.
However, unlike traditional RL, agentic RL demands substantial external cloud resources, e.g., CPUs for code execution and GPUs for reward models, that exist outside the primary training cluster.
Existing agentic RL framework typically rely on static over-provisioning, i.e., resources are often tied to long-lived trajectories or isolated by tasks, which leads to severe resource inefficiency. 

We propose the action-level orchestration, and incorporate it into \sysname, a unified resource management system that enables fine-grained external resource sharing and elasticity. 
\sysname utilizes a unified action-level formulation and an elastic scheduling algorithm to minimize action completion time (ACT) while satisfying heterogeneous resource constraints. Further, heterogeneous resource managers are tailored to efficiently support the action-level execution on resources with heterogeneous characteristics and topologies.
Evaluation on real-world agentic RL tasks demonstrates that \sysname improves average ACT by up to 4.3$\times$, speeds up the step duration of RL training by up to 1.5$\times$, and saves the external resources by up to 71.2$\%$. This system has been deployed to support the training of the MiMo series models.
\end{abstract}

\begin{document}

{
    \bgroup
    \setlength{\parindent}{0pt}
    \vspace*{3pt} 
    \begin{adjustwidth}{0pt}{0pt}  
    \begin{center} 
    {\titlefont ARL-Tangram: Unleash the Resource Efficiency in Agentic Reinforcement Learning \par}
    {
    \vskip5pt
    {\normalfont\sffamily\fontsize{11}{15}\selectfont Bangjun Xiao$^{\dagger\ddagger*}$ ~~~~~Yihao Zhao$^{\dagger\ddagger*}$ ~~~~~Xiangwei Deng$^{\dagger\ddagger}$ ~~~~~Shihua Yu$^{\ddagger}$ } \\
    {\normalfont\sffamily\fontsize{11}{15}\selectfont Yuxing Xiang$^{\dagger\ddagger}$  ~~~~~Huaqiu Liu$^{\ddagger}$ ~~~~~Qiying Wang$^{\ddagger}$ ~~~~~Liang Zhao$^{\ddagger}$ } \\
    {\normalfont\sffamily\fontsize{11}{15}\selectfont Hailin Zhang$^{\ddagger\S}$ ~~~~~Xuanzhe Liu$^{\dagger}$ ~~~~~Xin Jin$^{\dagger\diamond}$ ~~~~~Fuli Luo$^{\ddagger\diamond}$} \\
    \vskip10pt
    {\normalfont\sffamily\fontsize{11}{15}\selectfont $^{\dagger}$School of Computer Science, Peking University} \\
    {\normalfont\sffamily\fontsize{11}{15}\selectfont $^{\ddagger}$LLM-Core Xiaomi} \\
    % {\normalfont\sffamily\fontsize{11}{15}\selectfont $^{\S}$Independent Researcher}
    \vskip10pt
    }
    \end{center}
    \end{adjustwidth}
    \egroup
    {\abscontent}
    \thispagestyle{firststyle} 
}

\renewcommand{\thefootnote}{\fnsymbol{footnote}}
\footnotetext[0]{$^{*}$Equal contribution, work done during internship at Xiaomi Corporation.}
\footnotetext[0]{$^{\S}$Project leader.}
\footnotetext[0]{$^{\diamond}$Co-corresponding authors.}
\renewcommand{\thefootnote}{\arabic{footnote}}

\section{Introduction}

Agentic reinforcement learning (RL) has demonstrated significant performance improvements in handling complex real-world problems such as AI coding~\citep{le2022coderl,dou2024stepcoder}, deep search~\citep{zheng2025deepresearcher,qi2025webrl}, and embodied AI~\citep{zitkovich2023rt, wang2023voyager}.
Unlike traditional large language models (LLMs) that rely solely on internal knowledge, agents actively interact with real-world environments through external tools, such as shell commands, APIs, or device operations~\citep{cao2025skyrl}.
To enable these capabilities, agentic RL has become a critical component of the post-training pipeline and an important workload in cloud clusters~\citep{xiao2026mimo}.

\begin{figure}[t]

    \centerline{\includegraphics[width=0.9\linewidth]{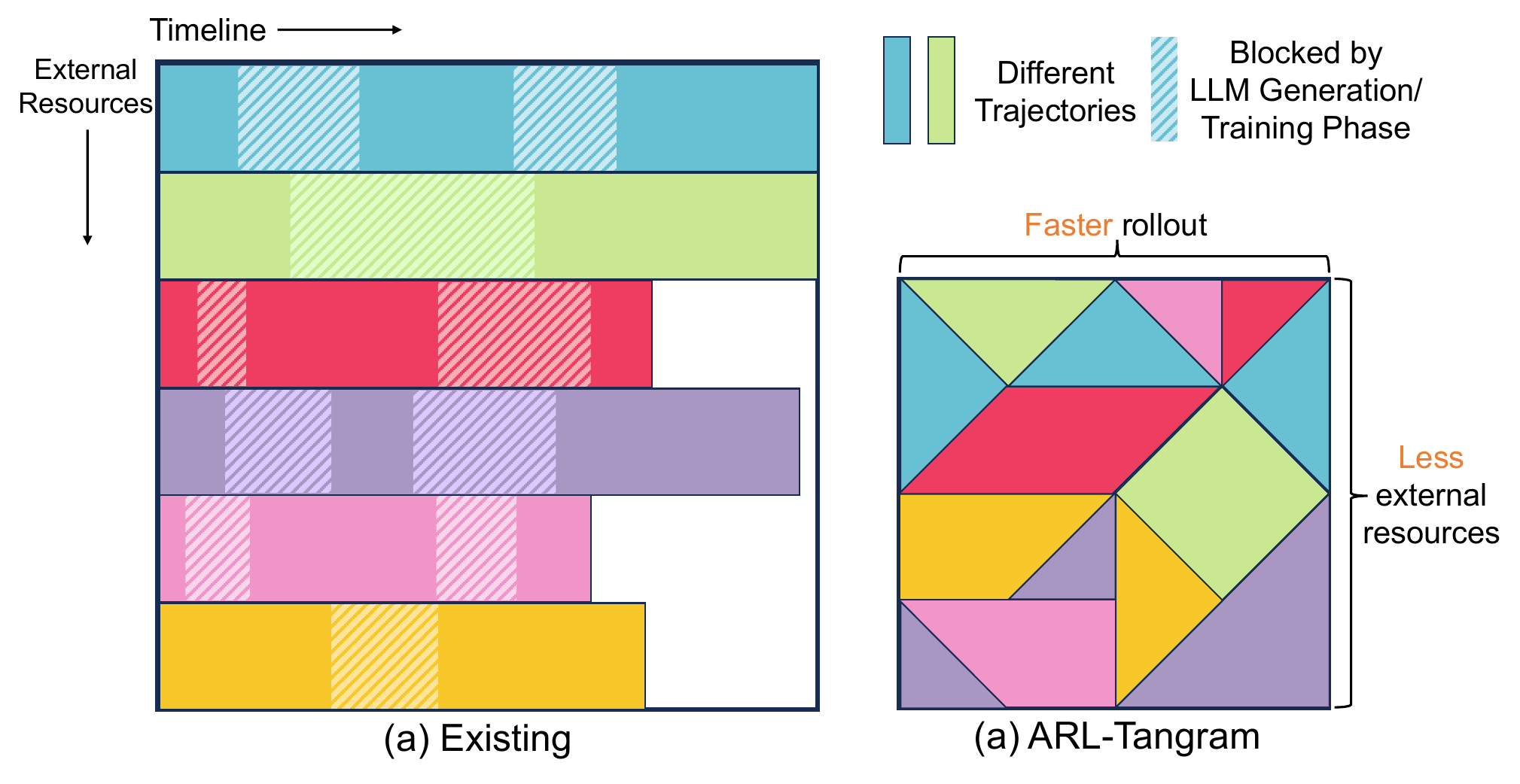}}
    \vspace{-0.2in}
    \caption{Comparison of existing approaches and \sysname. \protect\footnotemark}

    \label{fig:intro_op}
    \vspace{-0.2in}
\end{figure}

Unlike traditional RL, agentic RL demands substantial external resources, e.g., CPUs for AI coding, GPUs for reward models, and API quotas for web browsing.
These external resources, outside the primary training cluster, introduce additional management requirements.
Although this caters to the management approaches of cloud resources, the agentic scenario proposes higher demands for efficient management of these external resources, paramount for two reasons.
First, the latency of invocations is tightly coupled with the RL efficiency.
Without effective external resource management, invocations can become slow or even blocked, hindering the training speed, and even causing the entire training process to fail.
Second, the additional cost of external resources is non-negligible.
Agentic RL typically exhibits bursty invocation patterns, which requires provisioning numerous resources to maintain low invocation latency.
Thus, improving the utilization of these external resources and reducing the associated cost is undoubtably important.

\footnotetext{This is only a diagram for the insight of ARL-Tangram, where the patterns are not strictly triangles in practice.}

However, existing RL frameworks~\citep{sheng2025hybridflow,fu2025areal} often overlook the interdependencies between RL training and external resource usage.
These frameworks default to static resource-provisioning that leads to severe waste and inefficiency of external resources.
The over-provisioning exists at two levels (Figure~\ref{fig:intro_op}):
(1) At the trajectory level, systems often reserve dedicated resources, for the entire lifecycle of a trajectory, even though they are only invoked sporadically.
For example, in AI coding, the environment is accessed only for 47\% of its lifetime on average and leaving the allocated CPUs wasted for the remaining time.
(2) At the RL task level, tasks usually require specific external services and these services are deployed on isolated resources~\citep{jiang2025verltool,micromulti}.
However, due to the fluctuant pattern of external invocations, these resources are under-utilized and wasted.
In conclusion, previous static management limits system concurrency, leading to queuing delays and slower RL training.
% usually access different services and external resources are typically provisioned by services, preventing multiplexing even when sharing identical resource types.
% The phase alternation between rollout and training causes provisioned resources to sit idle during training phases.

% To address these inefficiencies, we propose \sysname, a unified action-level resource management system that decouples external resource management from the primary RL training cluster. 
% \sysname shifts the granularity of resource control from long-lived trajectories to individual atomic interactions, defined as "actions".
% By serving as a centralized intermediary between diverse RL tasks and a shared resource pool, \sysname enables statistical multiplexing across concurrent experiments and mitigates the resource idle time caused by long inactive agent environments.

% To address these inefficiencies, we propose the \textbf{action-level} scheduling that decouples external resource management from long-lived environments within trajectories or services within RL tasks.
% By \textbf{Breakdown}-ing the resource occupation of long-lived environments/services and \textbf{Pool}-ing the resources for invocations with the same resource type, we shift the granularity of resource control from original trajectory- or task-level into the level of atomic invocations, defined as "actions" in this paper. In addition, fine-grained resource management allows for elastic resource allocation to reduce the execution latency of actions.

To address these inefficiencies, we propose the \textbf{action-level} scheduling that shifts the granularity of external resource management from original trajectory- or task-level into a finer-grained action level, i.e., the level of atomic invocations.
We \textbf{breakdown} the resource occupation of long-lived environments/services and \textbf{pool} the resources for actions with the same resource type.
In addition, our fine-grained resource management allows for elastic resource allocation to reduce the execution latency of actions. 
As illustrated in Figure~\ref{fig:intro_op}, there are two RL tasks and 4 trajectories that invoke external resources of the same type, and the action-level scheduling enables less external resources by mitigating the over-provisioning and faster rollout by elastic resource allocation, compared with existing approaches.

However, realizing the action-level scheduling is non-trivial mainly because of three reasons. 
First, orchestrating actions with various external resources is complex.
Individual actions may exhibit requirements for multiple resource types, which is further complicated by varying elasticity and execution patterns of individual actions.
This necessitates a generalized abstraction model.
Second, the scheduler must operate subject to latency-sensitive workloads.
Action durations may be only microseconds, leaving an extremely short window for scheduling decisions.
This requires a lightweight algorithm capable of handling high-concurrency and bursty workloads.
Finally, how to uniformly and efficiently manage heterogeneous external resources with diverse characteristics and topologies is also challenging.
%action-level scheduling requires decoupling resource allocation from long-lived stateful agent environments.\zhl{what does this mean?}
% \sysname must release resources after each atomic action while preserving the environment state to ensure fast restoration, a direct conflict with traditional persistent allocation models.
% The system should be able to provide reliable resource sharing (which can release resources after each action while preserving the environment state), and low-overhead elasticity (which can deploy services with different parallelism fast).

Therefore, we design \sysname (\textbf{A}gentic \textbf{R}einforcement \textbf{L}earning \textbf{Tangram}), an action-level resource management system to uniformly orchestrate all these external resource invocations.
First, it utilizes a unified action formulation to manage actions with heterogeneous resource requirements and costs.
% heterogeneity, instead of treating actions as invocations that heterogeneous in costs and static in execution.
It formulates every action with a vectorized resource cost that accounts for various constraints, including CPU, GPU, memory, and API quotas.
Crucially, this formulation incorporates elasticity modeling, allowing the system to distinguish elastic actions and calculate the reduction in execution durations when allocating more resources.
This formulation enables \sysname to normalize various types of actions into a common format for scheduling.

The core of \sysname is an elastic resource scheduling algorithm designed to minimize the action completion time (ACT).
% \xbj{rewrite below}
Recognizing that shorter execution duration of actions improves the end-to-end efficiency of agentic RL training, 
we propose a heuristic scheduling algorithm, with a greedy eviction mechanism, to orchestrate actions based on the formulation and real-time system states. %, while maintaining low scheduling overhead.
The greedy eviction dynamically determines the scheduling strategies, avoiding overly aggressive/conservative allocation that lead to suboptimal ACT and degraded RL efficiency. %, while the entire scheduling algorithm is designed for actions with various invocation patterns.
% For practical usage, our scheduling algorithm maintains low overhead and is general for various action patterns.
% the scheduler dynamically performs resource allocation for each action, based on the formulation of actions and real-time system states.

% With these elastic scheduling decisions, 
\sysname tailors resource managers for heterogeneous resources, each with the specialized mechanism to efficiently \textbf{breakdown}, i.e., release the resources after each action and restore the states of environments or services when invoked, and effectively \textbf{pool}, i.e., allocate resources in the principle of mitigating fragmentation and improving parallel efficiency. 
Despite heterogeneous characteristics and topologies of resources, these managers expose a standardized interface to the scheduler, maintaining transparency of heterogeneous resources to the scheduling algorithm.
% The managers orchestrate actions of heterogeneous resources while maintaining the transparency in terms of the unified scheduler.

We implement \sysname as a standalone system regardless of the RL framework, external invocation types, and external resource types.
This design choice provides generality and simplicity to run with different external resources and across different RL frameworks.
We evaluate \sysname on real-world agentic RL training tasks, and the experimental results show that \sysname improves average ACT by up to 4.3$\times$, speeds up the step duration of RL training by up to 1.5$\times$, and saves external resources by up to 71.2$\%$. This system has been deployed to support the RL training of the MiMo series models.

In summary, this paper makes the following contributions,
\begin{itemize}[leftmargin=10pt]
% \vspace{-0.1in}
    \item We analyze the critical problem of external resource over-provisioning in agentic RL training, categorizing it into over-provisioning within trajectories and within RL tasks.
    \item We propose the action-level scheduling, and incorporates it into \sysname, a unified resource management system that shifts management from trajectory-level to action-level and enables fine-grained resource sharing and elasticity.
    \item We design a unified action formulation along with an elastic scheduling algorithm that minimizes ACT while respecting heterogeneous constraints. Besides, heterogeneous resource-specific managers further improve external resource utilization and training efficiency.
    \item We evaluate \sysname with real RL training tasks, showing that it significantly improves the end-to-end efficiency of agentic RL training and reduces external resource cost.

\end{itemize}

    % \item recognize and point out the critical issue during agentic RL training, over-provisioning of external resources
% \input{chapters/pre}
% \input{chapters/moediff}
\section{Background}

In this section, we first introduce the process of agentic RL training in \S\ref{sec:background_rl}.
Then we discuss the external resource management in \S\ref{sec:background_resource}, followed by the over-provisioning analysis in \S\ref{sec:background_op}.
Finally, we present our opportunities and challenges of external resource management in \S\ref{sec:background_opp}

\subsection{Agentic RL Training}
\label{sec:background_rl}
Agent shows significant performance improvement in handling real-world problems, including AI coding, deep searching, and embodied AI~\citep{dou2024stepcoder,qi2025webrl,wang2023voyager}.
Unlike traditional LLM, agentic LLM can interact with the real world through external tools, e.g., shell commands, APIs, or device operations.
The usage of external tools makes the RL pipeline of the agentic LLM different from the traditional LLM.
% Therefore, we need agentic RL as an important part during post-training.

\begin{figure}[t]
    \centerline{\includegraphics[width=0.7\linewidth]{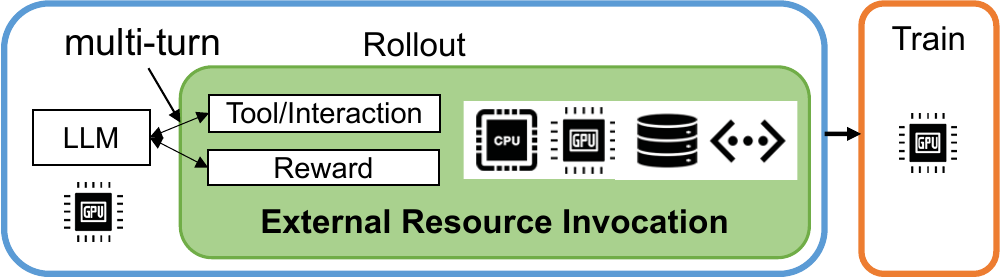}}
    \vspace{-0.1in}
    \caption{One training step of agentic RL.}
    \label{fig:background_workflow}
    \vspace{-0.2in}
\end{figure}

Agentic RL pipeline is illustrated in Figure~\ref{fig:background_workflow}. Given a batch of samples, the RL framework first uses an agentic LLM to process the samples and collects their execution trajectories, a phase known as \textit{rollout} in RL.
During the rollout phase, the framework typically follows the ReAct pattern~\citep{yao2022react}, 
where the LLM decides how to use tools based on the input prompt, then the RL framework accordingly performs tool invocations and gets the observation of the environment. 
Once a trajectory is completed, its reward is calculated based on either preference (e.g., RLHF~\citep{ouyang2022training} and LLM-as-a-judger~\citep{bai2022constitutional}) or ground-truth (e.g., RLVR~\citep{guo2025deepseek}), and subsequently used to train LLM.
% a loop of LLM generation and tool calling/environment interaction.
% First, the RL framework uses the LLM to decide how to use the tools based on the input prompt.
% Then, the RL framework performs tool invocations according to the LLM response and gets the observation of the environment.
% The observation is forwarded to the LLM for further decision.

Within a single trajectory, instead of invoking tools only once,  agentic RL usually repeats the ReAct pattern over multiple turns to enhance LLM's capacity of tool usage. 
While generation and training of LLM primarily consume GPU resources, tool invocations rely on heterogeneous resource types including CPU, GPU, storage, and network, termed as \textbf{external resources}  from the perspective of the RL framework.
Further, frequent invocations for external resources are a pronounced characteristic of agentic RL.
For example, AI coding repeatedly executes shell commands or edits files using CPUs, while DeepSearch accesses multiple websites and consumes API quotas.
In addition, reward computation at the end of trajectories typically involves external resources as well: LLM-as-a-judge is commonly deployed as an independent model service, and AI coding evaluates rewards by running a large amount of test cases.
All such operations that invoke resources outside the RL framework can be collectively referred to as \textbf{external resource invocations}. 

% Besides, reward services requiring GPUs to calculate the reward scores are becoming a trend~\citep{mopd,}

% The RL job generally falls into two categories: synchronous and asynchronous.
% For the synchronous RL, the training process occurs only after collecting the trajectories of every sample.
% The serial execution of synchronous RL introduces severe straggler bottlenecks: a single slow sample stalls the entire RL job.
% Conversely, asynchronous RL decouples the rollout phase and the training phase, allowing independent rollout phase while the training phase continuously consumes buffered collected samples.
% Asynchronous RL avoids the straggler problem, especially for agentic RL tasks, making it widely used in practice~\citep{}.

\subsection{External Resource Management}
\label{sec:background_resource}

% Although the demands for external resource invocations are urgent and general, there remain requirements for extra management efforts, because of three reasons.
Agentic RL training requires external resources to hold the environments, tools, or reward services.
The external resources require additional management efforts mainly due to two reasons, i.e., RL efficiency and resource cost.

% \xin{Add references or some experiment figures to support the three issues. You cannot just claim the issues without any evidence.}
% \xbj{TODO: add two figures, merge 1 and 3}

\begin{figure}[t]
    \begin{minipage}{0.24\linewidth}
        \centerline{\includegraphics[width=\linewidth]{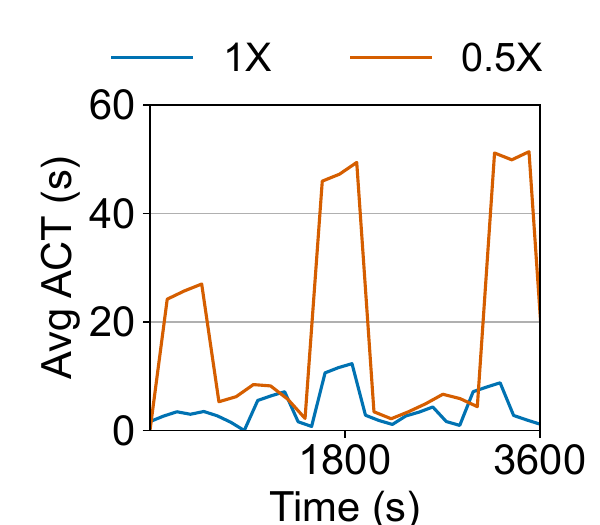}}
        % \vspace{-0.15in}
        \centerline{\small \quad(a)}
        % \vspace{-0.1in}
    \end{minipage}
    \begin{minipage}{0.24\linewidth}
        \centerline{\includegraphics[width=\linewidth]{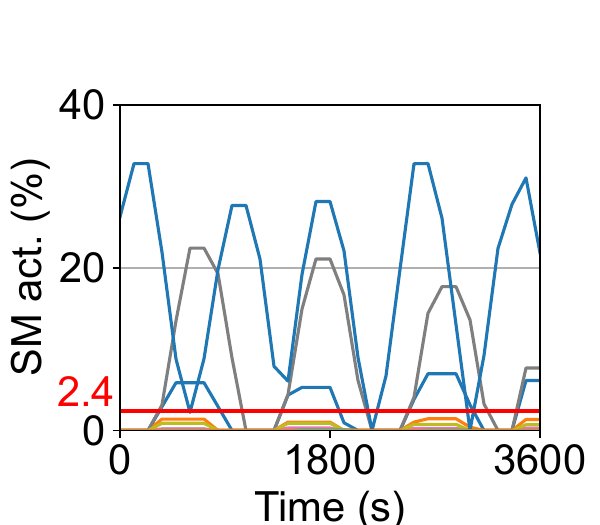}}
        % \vspace{-0.15in}
        \centerline{\small \quad(b)}
        % \vspace{-0.1in}
    \end{minipage}
    \begin{minipage}{0.24\linewidth}
        \centerline{\includegraphics[width=\linewidth]{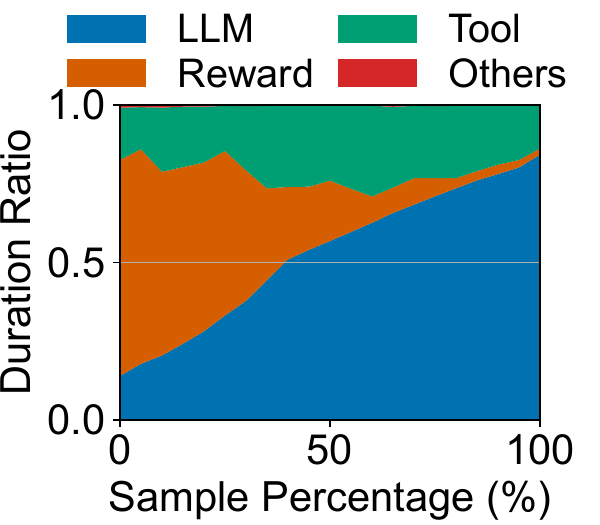}}
        % \vspace{-0.15in}
        \centerline{\small \quad(c)}
        % \vspace{-0.1in}
    \end{minipage}
    \begin{minipage}{0.24\linewidth}
        \centerline{\includegraphics[width=\linewidth]{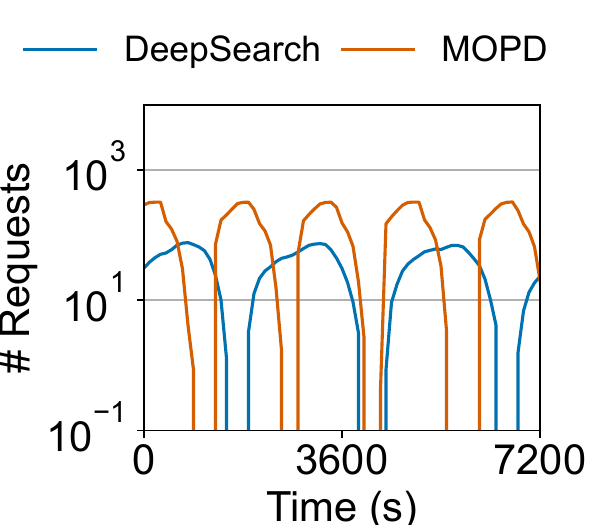}}
        % \vspace{-0.15in}
        \centerline{\small \quad(d)}
        % \vspace{-0.1in}
    \end{minipage}
    \vspace{-0.1in}
    \caption{(a): average ACT under 1$\times$ and $0.5\times$ external resource quantities. (b): SM activity of 12 different reward services in MOPD. (c): Code agent rollout duration ratio. (d): \# External invocations of two agentic RL tasks.}
    \label{fig:motiv_sm}
    \vspace{-0.2in}
    \end{figure}

First, the efficiency of RL tasks is highly related to how well the external resources are used.
Intuitively, if the external resource is insufficient or allocated inappropriately, the execution duration of external invocation is lengthened.
Note that, according to the commonly used agentic RL workflow in \S\ref{sec:background_rl}, the external invocations are all on the critical path of the rollout for each sample.
Thus, longer external invocations indicate longer rollout, and also longer RL training.
For example, we evaluate the same RL task under different amount of external resources.
Figure~\ref{fig:motiv_sm}(a) illustrates that the RL task with inappropriate external resources (0.5$\times$) has longer average ACT and step duration than the RL task with 1$\times$ resources.
More severely, if too many external invocations fail, the RL training task may also fail~\citep{jiang2025verltool}.

Second, the external invocations introduce additional resource cost, and it is highly motivated to reduce costs of external resources.
Usually, agentic RL requires deploying a CPU cluster for environment construction, or specialized GPUs that provide reward service for RL training, leading to extra expenses.
For example, a newly proposed distillation workflow, MOPD~\citep{xiao2026mimo}, uses dozens of teacher models with hundreds of GPUs to get the reward for one RL task.
% Approximately, MOPD leads to about xx\% additional GPU cost.
However, these external GPUs are not fully utilized.
Figure~\ref{fig:motiv_sm}(b) shows the SM activity of GPUs used by 12 teacher models.
The SM activity vibrates among time dimension and model dimension with an average less than 3\%,
which demonstrates great waste of external resources.

Existing frameworks~\citep{sheng2025hybridflow,fu2025areal,wang2025let} either overlook these interdependencies or default to static over-provisioning, leading to severe resource waste and cost inefficiency.

\subsection{Over-Provisioning of External Resources}
\label{sec:background_op}
% Existing works, along with some trivial approaches, 
The over-provisioning of external resources during agentic RL training can be categorized into two levels, both of which exacerbate training inefficiency and resource waste.

\parabf{Over-provisioning within trajectories.}
% Within a single trajectory, agents typically reserve or assume the availability of external resources (e.g., simulators, databases, search engines, or specialized inference services) for the entire duration of the rollout. However, actual interactions with these resources only occur at a small subset of time steps. As a result, resources remain idle for most of the trajectory execution, leading to low utilization and unnecessary blocking of other concurrent requests.
To guarantee the continuity and independence of multi-turn invocations, some agentic RL tasks require an isolated environment for each trajectory to interact with during the \textit{rollout} phase.
Existing frameworks~\citep{kubernetes}, statically allocate external resources for the entire lifecycle of a trajectory, although the trajectory only interacts with these resources within some time windows. 
We use a typical RL task in production, AI coding, to evaluate the action duration (including tool calling and reward calculation) of each trajectory.
As shown in Figure~\ref{fig:motiv_sm}(c), the time ratio of each trajectory is only 47\% on average. 

% Unlike the diagram, the distribution of LLM generation and environment invocation is not continuous. Instead, they are interleaved in the trajectory.
In practice, the LLM generation and external invocation phases are interleaved.
However, existing systems create environments and reserve external resources during the whole rollout phase.
% at the beginning of the trajectory and release them only after the trajectory terminates.
As a result, the reserved resources remain idle for most of the time, wasting resources that could otherwise be used by other trajectories. 
For example, the wasted external resources count for 53\% for AI coding (Figure~\ref{fig:motiv_sm}(c)).
% Consequently, a much larger CPU cluster is required to guarantee the stability and scale of agentic RL training, posing more severe challenges to both cluster scalability and cost efficiency.

Moreover, under finite external resources, over-provisioning within trajectories significantly limits system concurrency, increases queueing delays for tool invocations, and prevents allocating additional resource units to extremely computation-intensive operations.
These effects prolong rollout time and further reduce the overall efficiency of RL frameworks, as well as the utilization of GPUs in the training cluster.

% \begin{figure}[t]
%     \begin{minipage}{0.48\linewidth}
%         \centerline{\includegraphics[width=\linewidth]{figures/Motiv_traj_code.pdf}}
%         % \vspace{-0.15in}
%         % \centerline{\small (a).}
%         % \vspace{-0.1in}
%     \end{minipage}
%     \begin{minipage}{0.48\linewidth}
%         \centerline{\includegraphics[width=\linewidth]{figures/Motiv_req_counts.pdf}}
%         % \vspace{-0.15in}
%         % \centerline{\small (b).}
%         % \vspace{-0.1in}
%     \end{minipage}
%     \vspace{-0.1in}
%     \caption{Left: Code agent rollout duration ratio. Right: \# External invocations of two agentic RL tasks.}
%     \label{fig:motiv_op}
%     \vspace{-0.2in}
%     \end{figure}

\parabf{Over-provisioning within RL tasks.}
At the task level, different RL tasks usually require different external services to interact or calculate reward because of the service type and configuration~\citep{li2025cuda,jin2025search,xiao2026mimo}.
For example, different RL tasks that use reward models often send requests to separate reward services, because these models may differ in model architecture or parameters. 
These external services are usually deployed with different external resources~\citep{micromulti,jiang2025verltool}.

However, the external services are not fully used all the time, mainly because of the dynamism of external invocations along the entire RL training process.
For each RL task, there are even no external invocations during the training phase, leaving idle external services.
% Even during the rollout phase, the external invocations fluctuate considering the dynamism.
We profile the number of external invocations of two real agentic RL tasks, DeepSearch and MOPD, as shown in Figure~\ref{fig:motiv_sm}(d).
Note that the number of invocations varies up to three orders of magnitude, leading to severe waste of external resources.

% Apart from isolated environments for trajectories, agentic RL tasks also invoke various external services for consulting or judging.
% % The existence of over-provisioning within RL tasks is orthogonal to that within trajectories. In practice, 
% Therefore, they are usually provisioned with isolated clusters for external resource invocations, even when they rely on the same type of external resources.
% For example, different RL tasks that use reward models often send requests to separate reward services, because these models may differ in model architecture or parameters. 

% However, the utilization of external resources fluctuates significantly within a single RL task, in addition to within individual trajectories. Figure~\ref{fig:motiv_op}(b) illustrates the number of reward requests during the RL training of the DeepSearch Agent, and the number varies dramatically with the training proceeding. Due to the phase alternation between generation and training in many existing RL frameworks~\citep{sheng2025hybridflow,gao2025rollpacker}, interactions with external resources are blocked during the training phase. Consequently, the provisioned resources for this task remain idle for extended periods, resulting in severe resource waste.

\subsection{Opportunities and Challenges}
\label{sec:background_opp}
% According to the Agentic RL training pipeline described in section~\ref{}, each action to interact with external resources is on the critical path of \textit{rollout}. Meanwhile, during agentic RL training, the generation batch size on each data parallel (DP) group are relatively small compared with common RL training (usually 64 or even less, thus the inference engine fails to saturate the GPUs in the decoding stage), due to much longer lengths of trajectories. Therefore, once a trajectory begins to interact with external resources, the number of trajectories that are running in the inference engine decreases. Due to the existence of roof line model, the utilization of GPUs, i.e., SM utilization, decreases as well. 

\parabf{Opportunities in Action-level Scheduling.}
To address the inefficiency caused by the aforementioned over-provisioning, we propose to manage external resources at the \textbf{action-level}. 
Here, an \textit{action} refers to an atomic invocation of external resources, during which neither LLM generation nor any other actions in the trajectory are interleaved.

%implicitly 
Instead of allowing each trajectory to independently perform external invocations and reserve resources, all actions are submitted to a unified system that centrally manages heterogeneous external resources. 
In this system, invocations with the long-lived environments or services are uniformly scheduled and managed in the granularity of actions, i.e., \textbf{Breakdown} the resource occupation of long-lived environments/services. 
Further, the system uses pooled resources to serve these actions, and elastically allocates resources among them, i.e., \textbf{Pool} the resources. 
Through the process of \textbf{Breakdown} \& \textbf{Pool}, the over-provisioning problem is eliminated by the action-level scheduling, because the external resources are released or shrunk during the idle time.
% caused by LLM generation within trajectory or \textit{train} phases within RL tasks.
% Upon receiving an action request, the platform schedules its execution and elastically allocates reasonable resources based on the current system state (including resource availability, queue lengths, and contention levels) as well as the properties of the action (including predicted execution duration, resource type, and scalability). The execution result is then returned to the agentic RL framework and dispatched to the corresponding trajectory, enabling the alternating loop between LLM generation and tool invocation to proceed.

In addition to mitigating over-provisioning, action-level scheduling enables more flexible and fine-grained resource allocation, compared with the trajectory-level or task-level resource allocation in the past.
% The scheduler can assign the minimum required amount of resources tailored to each action, avoiding waste when the resource requirement of a certain action is extremely higher than others in a trajectory.
% Moreover, 
For example, when sufficient resources are available, it can dynamically allocate additional resource units to elastic actions, i.e., increase the degree-of-parallelism (DoP), to shorten execution durations, thereby further improving overall RL training efficiency.

In conclusion, by refining the granularity of resource management from trajectories or RL tasks to individual actions, action-level scheduling could:
(i) reduce resource idle time caused by long inactive periods through multiplexing across trajectories and tasks; and
% (ii) allocate resource units that are specifically matched to the requirements of each action; and
(ii) shorten interaction latency by dynamically scaling resource allocation.

\parabf{Challenges.}
% \xbj{exemplify heterogeneity? Additionally, although there is a unified abstraction modeling, heterogeneous resources and job patterns still challenge the generalizability. 
First, although action-level scheduling simplifies the workflow of individual actions, the diversity of resources and interaction patterns still poses difficulties in unified management (\S\ref{sec:design_modeling}). 
% For example, for some resources and job patterns, the execution duration is easy to predict, like the inference of reward models on GPUs. We can roughly predict the duration of a request through its input length along with the placement of targeted reward model. Instead, the extremely sophisticated scenarios on CPUs lead to the unpredictability of the execution durations for all CPU programs, among which only a small part of specific jobs with similar properties can be predicted in advance.
There are multiple external resources managed in the system, and an action may utilize more than one resource types. It is challenging to uniformly orchestrate these resources.
In addition, the patterns of actions, in terms of elasticity and execution durations, vary as well.
Therefore, designing a general abstraction, together with a set of unified interfaces to manage heterogeneous resources, is an essential foundation for building a unified scheduling algorithm.

Second, the scheduler must handle diverse and bursty workloads (\S\ref{sec:design_scheduling}).
Designing a unified and efficient scheduling algorithm under these conditions is challenging, since existing algorithms~\citep{grandl2016graphene,han2022microsecond,liu2024muxflow} are usually tailored to specific domains and do not generalize well to heterogeneous resources and job patterns.
Furthermore, the relatively short execution durations of actions (e.g., down to 1ms in AI coding) leave a quite short window for scheduling, making it impractical to apply heavyweight optimization techniques. %such as machine learning–based or convex optimization–based schedulers.

Third, enabling action-level scheduling requires decoupling resource allocation from long-lived stateful environments or services (\S\ref{sec:design_resource_managing}). Over-provisioning within trajectories/tasks originates from the need to maintain persistent state throughout an entire trajectory/task.
Therefore, a key challenge lies in how to release resources after each action while preserving the state of the environment/service and supporting real-time resource allocation and fast state restoration. 
% 2. efficient temporal multiplexing of external resources, reduce the overhead of multiplexing
In addition, it is non-trivial to support elastic DoP and dynamic resource scaling for heterogeneous resources with diverse characteristics and topologies.

% Finally, many existing approaches focus primarily on job reordering~\citep{} rather than elastic resource allocation, while both overly aggressive and overly conservative scaling strategies can lead to suboptimal interaction latency and degraded training efficiency.

\section{Architecture}

\begin{figure}[t]
    \centerline{\includegraphics[width=0.6\linewidth]{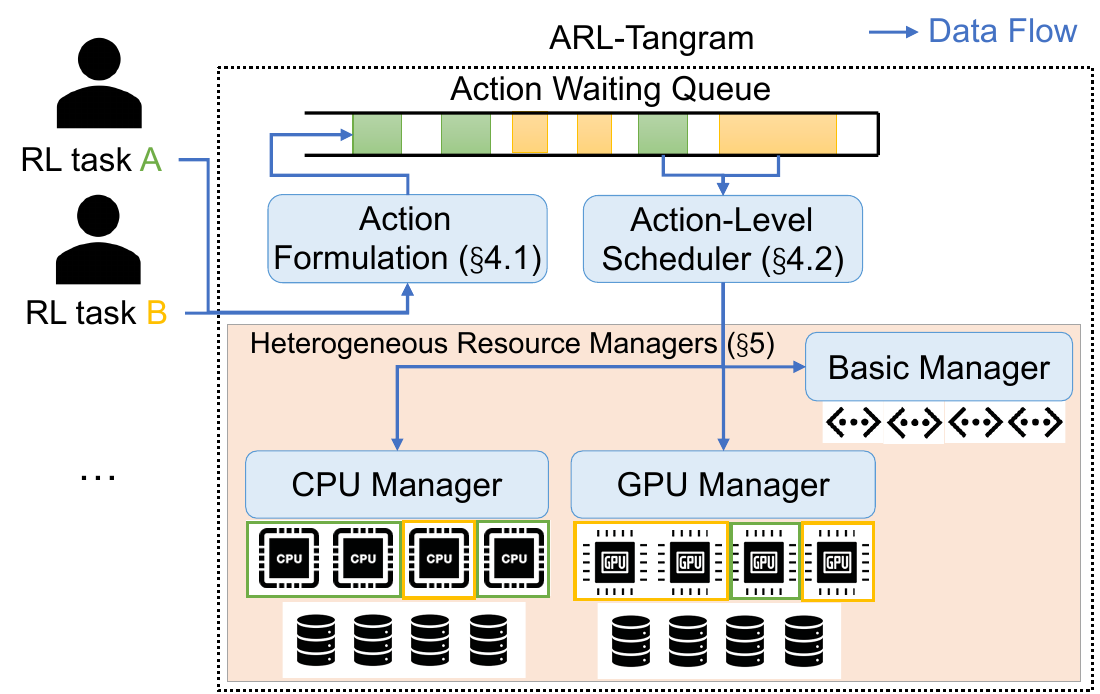}}
    \vspace{-0.1in}
    \caption{System overview of \sysname.}
    \label{fig:arch_arch}
    \vspace{-0.15in}
\end{figure}

To address the multifaceted over-provisioning issues inherent in current agentic RL training—within trajectories, and within RL tasks. 
We propose a unified action-level platform, \sysname. 
Our architecture decouples external resource management from the primary RL training cluster, shifting the granularity of control from long-lived trajectories to individual atomic interactions, or actions.

As shown in Figure~\ref{fig:arch_arch}, the system serves as a centralized intermediary between diverse RL Tasks and heterogeneous external resources. 
The workflow follows a standardized execution cycle:
(1) Action Submission: During the rollout phase, when an agentic LLM invokes a tool, or reward calculation with external resources, it submits a request to \sysname.
(2) Unified Formulation \& Queuing: Through formulation, \sysname translates various external resource invocations into unified actions. Hence, these requests are submitted to a Unified Action Queue.
(3) Elastic Scheduling: The scheduler schedules the actions in the waiting queue under the constraints of external resources, and coordinates with specialized Resource Managers (e.g., CPU, GPU, API) to elastically allocate the resources based on real-time system state and action formulation.
(4) Action Execution: After the units of resources are decided, actions are independently executed on the resources allocated by heterogeneous resource managers.
(5) Transmit \& Observation: Once an execution is complete, the results are transmitted back to the RL framework, allowing it to proceed with the next LLM generation.

There are three main components in \sysname---the action formulation module(\S\ref{sec:design_modeling}), the action-level scheduler(\S\ref{sec:design_scheduling}), and heterogeneous resource managers(\S\ref{sec:design_resource_managing}).
The foundation of our architecture treats every atomic invocation as an action.
Thus, the formulation module models the costs and performance elasticity of various actions in a unified way.
Then, the scheduler decides whether an action is able to schedule under the resource constraints, and allocates resources elastically to minimize the action completion time (ACT).
In this way, \sysname reduces the time LLMs spend waiting for external resource invocations, further accelerating the RL training.
With scheduling decisions, heterogeneous resource managers are responsible for efficient resource allocation and sharing.
Using a unified resource interface, our resource managers, including Basic Manager, CPU Manager and GPU Manager, employ domain-specific techniques to improve system robustness, resource utilization, and action speed.

\section{Unified Action-Level Formulation and Scheduling}

We first formulate the entire workflow at the action level in \S\ref{sec:design_modeling}.
Our formulation unifies heterogeneous external resource types and diverse tool types.
Then we introduce our elastic resource scheduling algorithm in \S\ref{sec:design_scheduling}.
% Several primitives are defined to acquire and interpret these arguments.

\subsection{Action Formulation}
\label{sec:design_modeling}

\paraf{Vectorized Resource Cost Modeling.} 
At first, to enable unified action-level scheduling, we need a clear and unified representation of the resource cost.

Given an action, $a_i$, it may consume multiple types of resource simultaneously. For example, an action of the DeepSearch RL task may involve parallel or subsequent queries to multiple websites, where each website has a unique API quota and is considered as a different resource type.
An action of the AI coding task involves not only CPUs but also memory.
Thus, the cost of an action is represented as a multi-dimensional vector, denoted as $C_i = (c_{i,0}, c_{i,1},...,c_{i,k-1})$, where $k$ refers to the number of resource types managed by \sysname.
The vector enables joint consideration of multiple resource constraints during scheduling and allocation.
Each dimension of the vector corresponds to the consumption of one type of resource managed by \sysname (including CPU, GPU, memory, GPU memory, network bandwidth, and external API quotas).

Importantly, the cost for each resource type is not a static value.
For actions that are elastic to the resource quantity, the cost is specified as a range between the minimum and maximum feasible resource usage.
Even more sophisticated, for some specific tasks like reward services occupying multiple GPUs, the allowed unit of resource is discrete (e.g. 1, 2, 4, 8).
Therefore, the $c_{i,j}$ in $C_i$ has a specific constraint, representing its all possible resource quantity.
This formulation allows the scheduler to flexibly adjust resource allocation without violating feasibility constraints or introduce some priors to narrow the search space during scheduling.

\parabf{Elasticity Modeling.} For elastic actions, an additional argument is introduced to describe their elastic behavior.
In terms of a specific resource, its elasticity is described by a mapping from the $m$ units of the resources to a elastic ratio between 0 and 1. And the function satisfies the following:
\begin{equation}
\label{equ:scale}
\begin{aligned}
% a.\text{getDur}(m) = \frac{T_{ori} - T_{invar}}{S(m) * m} + T_{invar},
a.\text{getDur}(m) = \frac{T_{ori}}{E(m) * m}\quad (0<E(m)\le1),
\end{aligned}
\end{equation}
where $T_{ori}$ refers to the execution duration with a single unit.
%and $T_{invar}$ refers to the part of original duration that is invariant to the allocated units.

To simplify modeling and optimization, we assume that among all resource types consumed by an action, only one resource type is the key elasticity resource. 
This assumption is reasonable in many practical scenarios.
For instance, GPU-based actions require sufficient GPU memory for execution, but their execution latency is primarily determined by the degree of parallelism.
Under this assumption, the elasticity of an action is defined as a mapping from the usage of the key resource type, $S(m)~( m \in c_{i, key},~0\le key < k)$, which quantifies how the execution duration changes with the allocated amount of the key resource.

\parabf{Execution Duration Modeling.}
To prepare for the elastic scheduling algorithm, it is essential to incorporate execution durations into the action specification in addition to elasticity.
Therefore, we introduce the original execution duration and normalize it with a single unit, as aforementioned in Eq.\ref{equ:scale}, for actions whose execution duration is profileable. 
These profiles, along with the definition of elasticity, allow the system to estimate the time cost under different resource allocations, which is essential for optimizing scheduling decisions and balancing performance with resource efficiency.

\subsection{Elastic Resource Scheduling}
\label{sec:design_scheduling}

An elastic resource scheduling algorithm for actions, motivated by the action-level resource allocation and scheduling, is required to guarantee the efficiency under finite resources. 

We first formulate the action completion time (ACT) as the optimization objective of the scheduling algorithm. 
Analogous to the job completion time (JCT), ACT is decomposed into: (1) the duration when the action queuing to be scheduled; (2) the execution duration of the action. The scheduling objective is therefore to find the resource allocation among several actions that minimizes the sum of ACTs:
\begin{equation}
\label{equ:obj}
\begin{aligned}
 \text{Objective:} \quad {\text{minimize}} ~ ACTs =\sum_{0\leq i < N} (T^q_{i} + T_i),
\end{aligned}
\end{equation}
where $N$ denotes the number of actions in the waiting queue at the scheduling time, $T_i$ denotes the actual execution duration of $ith$ action, and $T^q_{i}$ denotes its queuing duration.

Specific to the elastic scheduling algorithm, the problem consists of two coupled problems: (1) determining the execution order of actions in the waiting queue, and (2) determining how many units of resources to allocate for each action, both of which are non-trivial. 
To narrow the search space and align with the key observation of this paper, resource over-provisioning, we mainly focus on the second problem. %, while adopting a trivial algorithm for the first is enough. 
Since all actions lie on the critical path of their corresponding trajectories, agentic RL training is highly sensitive to the issue of starvation (a starved action may invalidate the entire trajectory). Therefore, we choose a First-Come First-Served (FCFS) policy to determine the scheduling order.
Our scheduling algorithm can work with other ordering algorithms.

The second problem is a classical problem: allocating finite units of resources among several elastic jobs and minimizing the total JCTs, which is NP-hard~\citep{bruno1974scheduling}. As discussed in \S~\ref{sec:background_op}, the quite short window of scheduling necessitates a heuristic solution, and the heterogeneity of both resource types and job patterns requires generalizability. Accordingly, we propose a unified heuristic scheduling algorithm, primarily based on a greedy eviction algorithm with approximated objective, as shown in Algorithm~\ref{alg:ersa}.

\begin{algorithm}[t!]
\captionof{algorithm}{Elastic Resource Scheduling Algorithm.}\label{alg:ersa}
    \begin{footnotesize}
    \begin{algorithmic}[1]
        \Require Actions in the waiting queue $W$, vectorized resource status $R = (R_0, ..., R_{k-1})$
        \State $Selected \gets \varnothing$
        \State $C \gets W[:n],~n=\max\{i \mid R.\text{accommodate}(W[:i])\}$ \algcomment{Get candidates}
        \For{$j = 0 \rightarrow k-1 $} \algcomment{Process actions sharing key elasticity resource $R_j$}
            \State $C_j \gets C.\text{split}(R_j)$
            \If{ $\forall a_{i} \in C_j$: ($E_{i}$ is None or $S_{i} \equiv 0$)} \algcomment{if elasticity unknown}
            \State $Selected+= C_j$;~\textbf{continue} \algcomment{or zero, directly continue}
            \EndIf
            %\textcircled{3}:scalability和对应的execution duration均已知
            \State $obj \gets \text{getApproximatedObjective}(C_j, R_j)$ \algcomment{Initialize objectives}
            \For{$t = 1 \rightarrow |C_j|$} \algcomment{Greedily evict the last action and compare}
            \State $newObj \gets \text{getApproximatedObjective}(C_j[:-t], R_j)$
            \If{$newObj >= obj$} \textbf{break}
            \Else~~$obj \gets newObj$
            \EndIf
            \EndFor
            \State $Selected += C_j[:-t+1]$;~\textbf{continue} 
        \EndFor
        \Return $Selected$
    \end{algorithmic}
    \end{footnotesize}
\end{algorithm}

Each time the scheduling algorithm is invoked, we first select the first $n$ actions from the waiting queue as candidates $C$ (due to FCFS), ensuring that vectorized resource constraints $R = (R_0, ..., R_{k-1})$ are satisfied with each action allocated with least-required resource units, i.e.,
$$
R_j \ge \{c_{0,j}^{min}, ...,c_{n-1,j}^{min}\},
$$
where $\ge$ indicates that not only the remaining units of $j$th resource type is no smaller than the sum of the minimum requirements $c_{i,j}^{min}$, but also the current status in terms of its actual topology can accommodate all these candidates (Line 1-2).
We then split the candidates into several groups according to the key elasticity resource (\S~\ref{sec:design_modeling}), and respectively apply the elastic allocation algorithm on each resource type. Granted by the assumption that an action’s elasticity is dominated by a single resource type, scaling an action along this resource does not affect the allocation of other resources (Line 3-4).
For each resource type, the algorithm selects a subset of candidates and determines their allocated units.

For scalable actions that both elasticity and execution durations are known, we propose a greedy algorithm by approximating the objectives, ACTs, during the decision process. First, we schedule all candidates $C_j$ using their minimum resource requirements and compute the initial objective.
Second, we iteratively evict the last action from candidates $C_j$ and reassign its resources to the remaining actions. For each reduced candidate set $C_j[:t]$, we compute the optimal resource allocation and the corresponding objective. Once further eviction fails to reduce the objective, which means that evicting this action and scaling the left candidates fails to optimize the ACTs, the greedy algorithm terminates. 

As for objective approximation (the pseudo code is shown in Appendix), we decompose it into two parts, i.e., the ACTs of the candidates and the ACTs of the remaining actions in the waiting queue. Because the candidates are to be scheduled at once, their ACTs are computed exactly using their determined allocations. We employ a topology-agnostic dynamic programming (DP) algorithm that is applicable to heterogeneous resource topologies, termed as DPArrange in the pseudo code. This algorithm, provided in Appendix~\ref{app:udp}, resolves the optimal discrete allocation among these scalable candidates and generates the updated completion heap according to the allocation.
% , while the topology-aware variant is introduced in following \S~\ref{}.
However, ACTs of the remaining actions are influenced by the allocation strategy ahead of them. It is intractable to recursively compute the best allocation and attain the optimal objective for these actions. Instead, we simplify the process and estimate their ACTs by assigning minimum resources and sequentially inserting them into the completion heap.

Notably, to approximate the objective more precisely, we introduce a parameter $depth$, which expands the search space by allowing the first remaining action to explore multiple allocation choices. In practice, setting $depth=2$ or $3$ is sufficient. Moreover, the conditions require known execution durations only for scalable actions, rather than all actions. This is because, during greedy eviction of candidates, the durations of non-scalable actions remain constant, and do not affect the relative comparison of objectives.
Although their exact execution durations affect the completion heap, it is acceptable to be approximated by historical averages, due to the fact that scalable actions typically last for much longer duration than non-scalable ones and dominate the evolution of the completion heap. The computational complexity of this algorithm is dominated by the implementation of the DP algorithm. Using the classical implementation for common topologies, the overall complexity is $O(kn^2m^2)$, where $n$ refers to the number of candidates and $m$ refers to the remaining units of a resource type.

As for the non-scalable resource or the actions whose elasticity are unknown, the algorithm directly select these candidates and allocate them with least-required units, while guaranteeing the vectorized resource constraints satisfied.

\section{Heterogeneous Resource Managers}
\label{sec:design_resource_managing}

While providing a unified abstraction and scheduling algorithm, achieving \textbf{Breakdown} \& \textbf{Pool} (\S\ref{sec:background_opp}) additionally requires resource-specific management policies. 
In particular, \textbf{Breakdown} focuses on releasing resources while preserving execution states or enabling fast restoration, i.e., efficient context switching, as well as supporting elastic degrees of parallelism for each action.
\textbf{Pool} emphasizes the design of resource allocation that mitigate fragmentation and improve parallel efficiency, thereby backing up the unified elastic scheduling framework.
As discussed in Section~\ref{sec:background_opp}, heterogeneous resources differ significantly in both characteristics and topologies.
We therefore design specialized resource managers for different resource types, including CPUs, GPUs, and other basic resources.
%, and present their design principles.

\subsection{Basic Resource Manager}
For external resources that cannot be scaled up, such as website quotas and request QPS limits, we introduce a \emph{Basic Resource Manager} to prevent contention and violations of resource constraints.
This manager supports two consumption patterns: concurrency-based (limiting the maximum concurrent usage) and quota-based (limiting the total usage in a period).
It also defines a set of basic interfaces that can be inherited and selectively overridden to support action-level scheduling and unified resource management.

\subsection{CPU Manager via AOE}

In current cloud CPU clusters, the combination of Kubernetes (k8s) and Docker is the de facto standard, where Docker provides environment isolation and k8s performs cluster-level resource management. However, designed for long-running cloud services, k8s incurs substantial scheduling overhead 
and only manages resources at the granularity of pods. Resource quotas are fixed at pod creation time\footnote{In-place resizing is not yet available in stable releases and introduces unacceptable latency even in nightly versions.}, which leads to over-provisioning within trajectories in agentic RL workloads.
To address this limitation, we introduce a CPU management mechanism termed \emph{allocate-on-execution (AOE)}, which directly interacts with containers and prepares for fine-grained CPU allocation.

\parabf{Breakdown in CPU Manager.} 
AOE leverages Docker’s \texttt{update} interface to modify a container’s cgroup configuration on demand, including parameters like \texttt{cpulimit} and \texttt{cpuset}. Before each invocation of \texttt{docker.exec()}, CPU Manager updates the container’s cgroup according to the resource units assigned by the scheduling algorithm. Docker then forks a process under the updated cgroup to execute the action. After execution, the process terminates and the allocated CPU resources are reclaimed.
Notably, to preserve the long-lived state of each environment, the memory allocated to each container is preserved, which is acceptable in modern clusters with abundant memory capacity.

Adjusting cgroup configuration alone is insufficient to achieve elastic DoP, since not all elastic actions innately are allowed for parallel in actual. Therefore, we should ensure the elasticity of certain actions when constructing RL datasets or submitting actions. 
% \zhl{is elasticity suitable for ``Breakdown''? maybe better in ``Pool'', or a new primitive? }
For example, in code generation tasks, test cases can be efficiently executed in parallel, yet the testing commands provided in existing datasets typically support only single-process execution. As most test commands rely on the library \texttt{pytest}, we replace \texttt{pytest} with \texttt{pytest -n \$\{parallel\_num\}} and dynamically set \texttt{parallel\_num} at runtime to enable parallel execution.

\parabf{Pool in CPU Manager.}
The design of resource allocation in CPU clusters is relatively straightforward. Since CPU cores and CPU memory are tightly coupled, CPU Manager jointly manages these two resource types. For CPU allocation, the manager not only determines the number of cores assigned to each action but also explicitly specifies the set of cores used for parallel execution. Each core is exclusively occupied by a single action at any time, preventing interference across actions.
In addition, parallel efficiency is affected by inter-core communication overhead, which depends on the physical distances between cores on the chip. Therefore, when allocating CPU cores for elastic actions, CPU Manager prioritizes selecting cores within the same Non-Uniform Memory Access (NUMA) node.

When the first action of a trajectory is invoked, CPU Manager filters nodes that have sufficient resources—namely, enough CPU cores for the action and sufficient memory for the entire trajectory—and selects one node according to a CPU memory load-balancing policy. All subsequent actions within the same trajectory are then constrained to execute on the selected node.
Although CPUs across different nodes cannot cooperate for parallel execution, modern cloud CPU nodes typically provide 128 cores or more. This capacity is large relative to the number of CPU cores allocated per action, and as a result, fragmentation of CPU cores is generally not severe in practice. Therefore, CPU Manager independently performs the scheduling algorithms for each node.

\begin{figure}[t]
    \centerline{\includegraphics[width=0.8\linewidth]{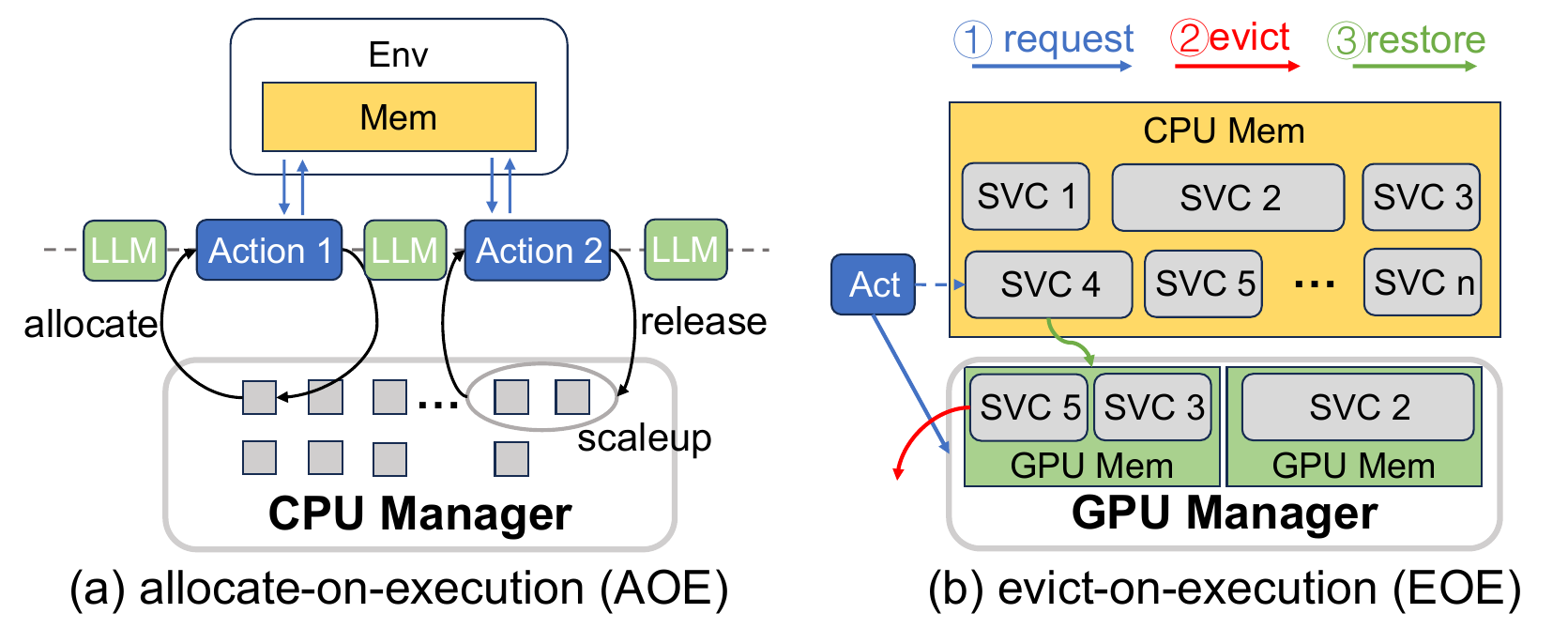}}
    \vspace{-0.1in}
    \caption{Heterogeneous resource management.}
    \label{fig:design_manager}
    \vspace{-0.15in}
    \end{figure}

\subsection{GPU Manager via EOE}
Compared with Docker-based CPU environments, long-lived services on GPU clusters incur significantly higher setup overhead. For example, a reward model service must compile kernels, establish communication groups, and, most critically, load model parameters into GPU memory~\citep{zhang2025blitzscale,lou2025hydraserveminimizingcoldstart}.
Moreover, GPU memory is both scarce and heavily contended, preventing all services from being persistently cached. Relying solely on CPU-memory caching would introduce prohibitive restoration overhead. We adopt an \emph{evict-on-execution (EOE)} mechanism to reduce context-switch overhead, and design corresponding resource allocation strategies to mitigate resource fragmentation and service dithering.

\parabf{Breakdown in GPU Manager.} 
Upon initialization, GPU Manager iteratively prepares all required services by deploying them on each feasible group of GPUs and backing up their states in CPU memory. When an action requests a service, GPU Manager allocates a set of GPUs with required units and checks whether the requested service is already resident in GPU memory. If so, the action is executed immediately.
Otherwise, the GPU Manager restores the service from CPU memory, evicting cached services from GPU memory as necessary until sufficient memory is available for exclusive execution. After the action completes, the restored service remains cached in GPU memory until it is evicted at a later time. 
Although eviction and restoration inevitably introduce additional latency, we observe that GPU memory states of many services remain unchanged across invocations. As a result, eviction does not require writing service states back to CPU memory; it is sufficient to release the occupied GPU memory while keeping an invariant copy in CPU memory.
Regarding the overhead of restoring services from CPU memory, prior work~\citep{zhang2025blitzscale,xiang2025aegaeon} has shown that this cost can be effectively reduced, making the overhead acceptable in practice.

Elastic DoP is naturally supported under EOE by treating different DoP configurations of a service as distinct services. An action requesting a service can be routed to any instance of arbitrary DoP, and relies on the scheduling algorithm to make decision.
Notably, EOE guarantees that all cached services have sufficient GPU memory to serve actions, while GPU allocation enforces that at most one action executes on each GPU at any time. This design enables a GPU cluster to host multiple services efficiently for agentic RL training.

\parabf{Pool in GPU Manager.}
The specialized topology of GPU clusters—characterized by fewer devices per node and strong demands for parallel execution—necessitates careful mitigation of GPU fragmentation.
We adopt a multi-level cell structure~\citep{zhao2020hived} to organize and manage GPU resources.
Targeting common DoPs, a legal allocation set, termed \emph{chunk}, is defined as a contiguous GPU interval $(start, end)$ satisfying:
$$
end - start = 2^a~~ \text{and}~~ start | 2^a\quad (a \in \{0,1,2,3\}),
$$
where $a$ refers to the level of this chunk.
Initially, each GPU node is represented as a single available chunk $(0, 8)$ with level 3. When allocating $m$ GPUs, where $2^{a-1} < m \le 2^a$, GPU Manager traverses available chunks in the order of levels and selects the smallest chunk with level $b~(b\ge a)$. If $b > a$, GPU Manager splits the chunk into several legal chunks accordingly, and the resulting chunk is returned.
In addition, GPU Manager applies a least-recently-used (LRU) eviction policy, to reduce dithering of service cache. When multiple chunks of the same level are available, preference is given to the chunk that already caches the requested service.

All interfaces required by the unified scheduling algorithm and related to the topology of external resources are overloaded in GPU Manager, particularly the topology-agnostic DP algorithm, which is detailed in Appendix~\ref{app:udp}.
\section{Evaluation}

We first evaluate \sysname using various agentic RL tasks to show its ACT efficiency (\S\ref{sec:e2e}),
and then highlight its scalability over RL batch size and external resource (\S\ref{sec:eval_scale}).
Finally, we investigate the design of \sysname and its overhead (\S\ref{sec:eval_analyse})

\begin{figure*}[t]
    \begin{minipage}{0.24\linewidth}
        \centerline{\includegraphics[width=\linewidth,trim=0 0 0 20,clip]{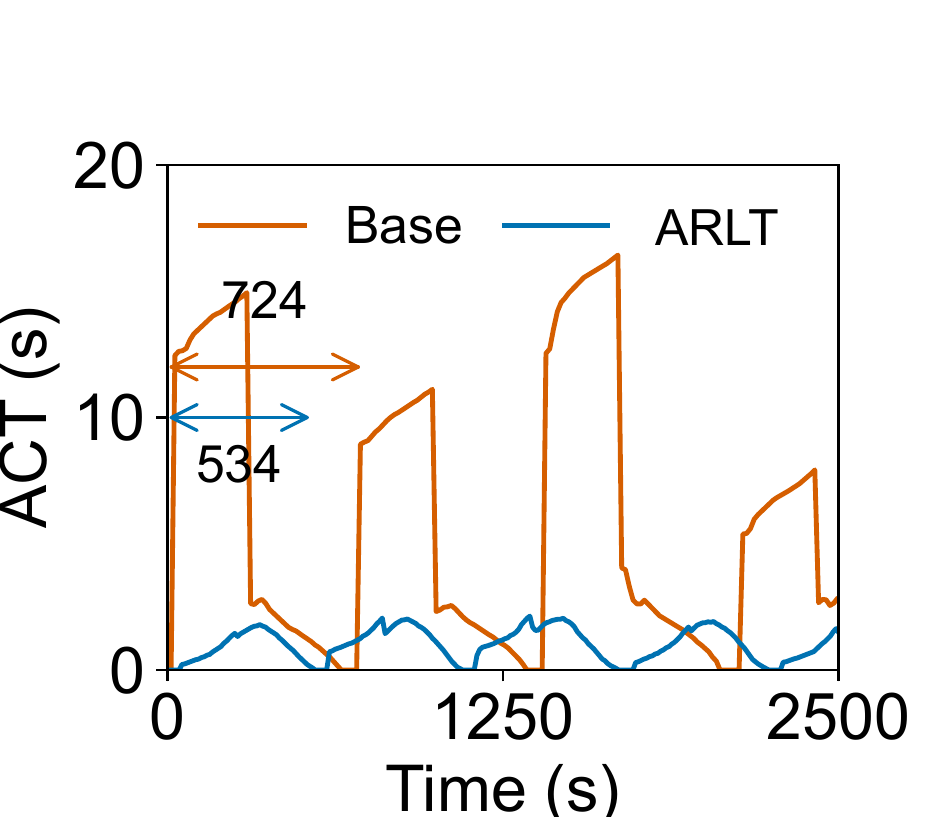}}
        % \vspace{-0.15in}
        \centerline{\small (a) AI Coding.}
        \vspace{-0.1in}
    \end{minipage}
    \begin{minipage}{0.24\linewidth}
        \centerline{\includegraphics[width=\linewidth,trim=0 0 0 20,clip]{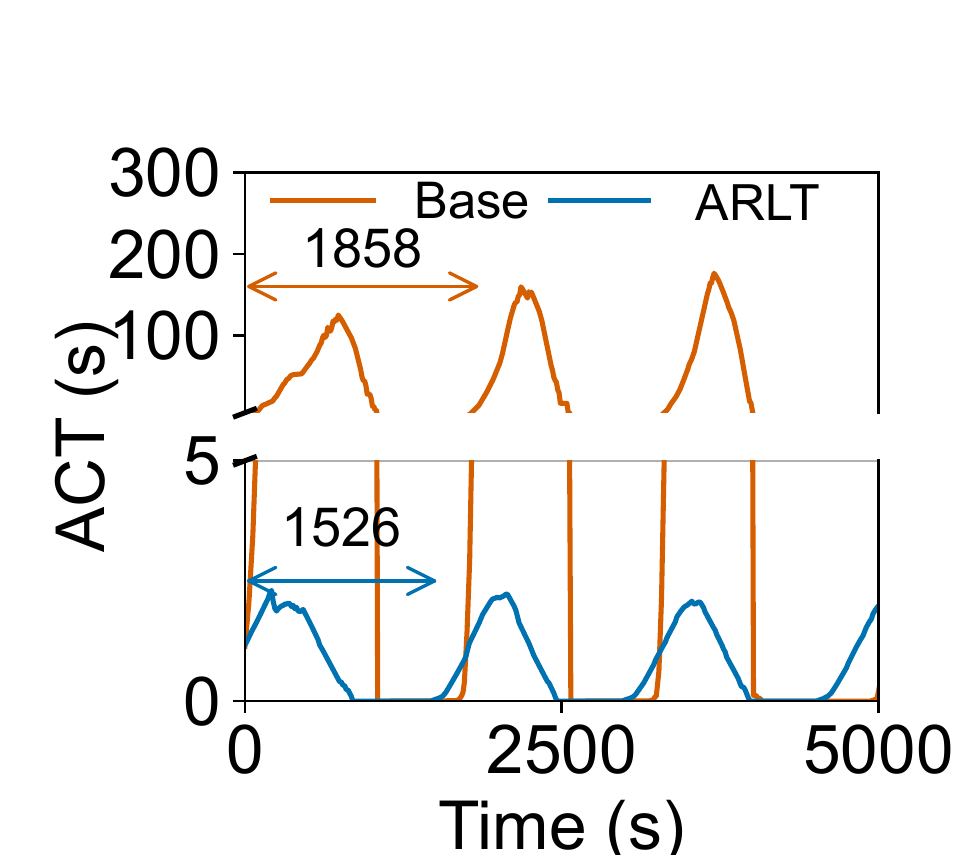}}
        % \vspace{-0.15in}
        \centerline{\small (b) MOPD.}
        \vspace{-0.1in}
    \end{minipage}
    \begin{minipage}{0.24\linewidth}
        \centerline{\includegraphics[width=\linewidth,trim=0 0 0 20,clip]{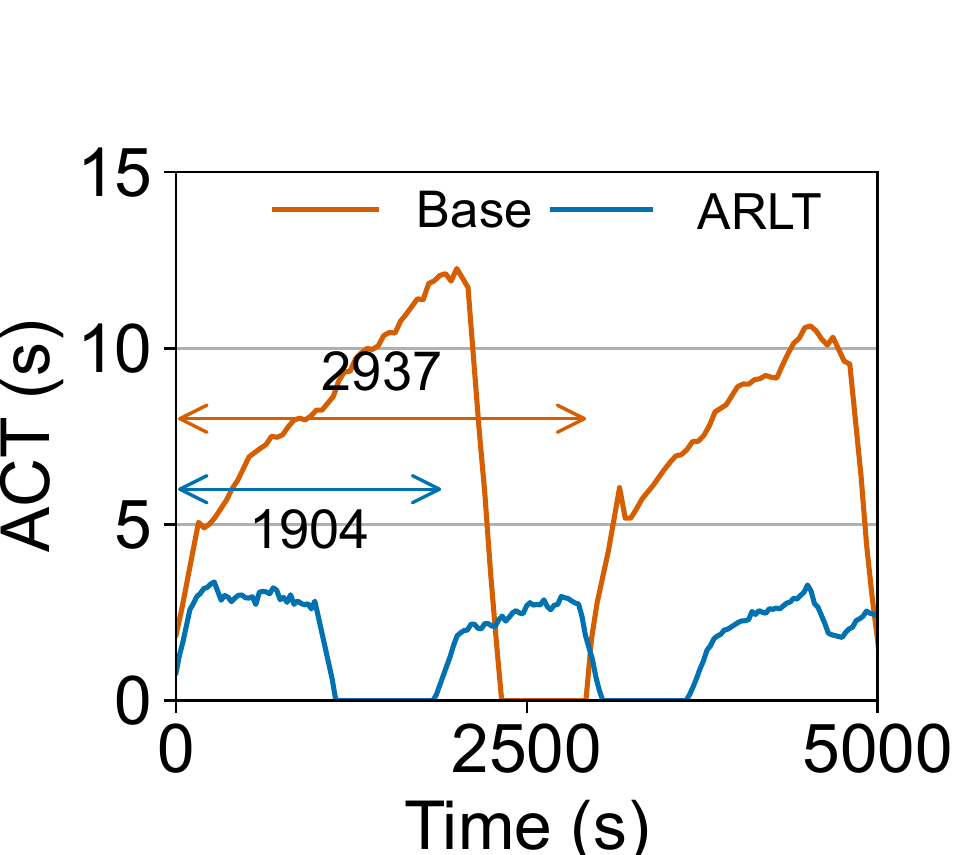}}
        % \vspace{-0.15in}
        \centerline{\small (c) DeepSearch.}
        \vspace{-0.1in}
    \end{minipage}
    \begin{minipage}{0.24\linewidth}
        \centerline{\includegraphics[width=\linewidth,trim=0 0 0 20,clip]{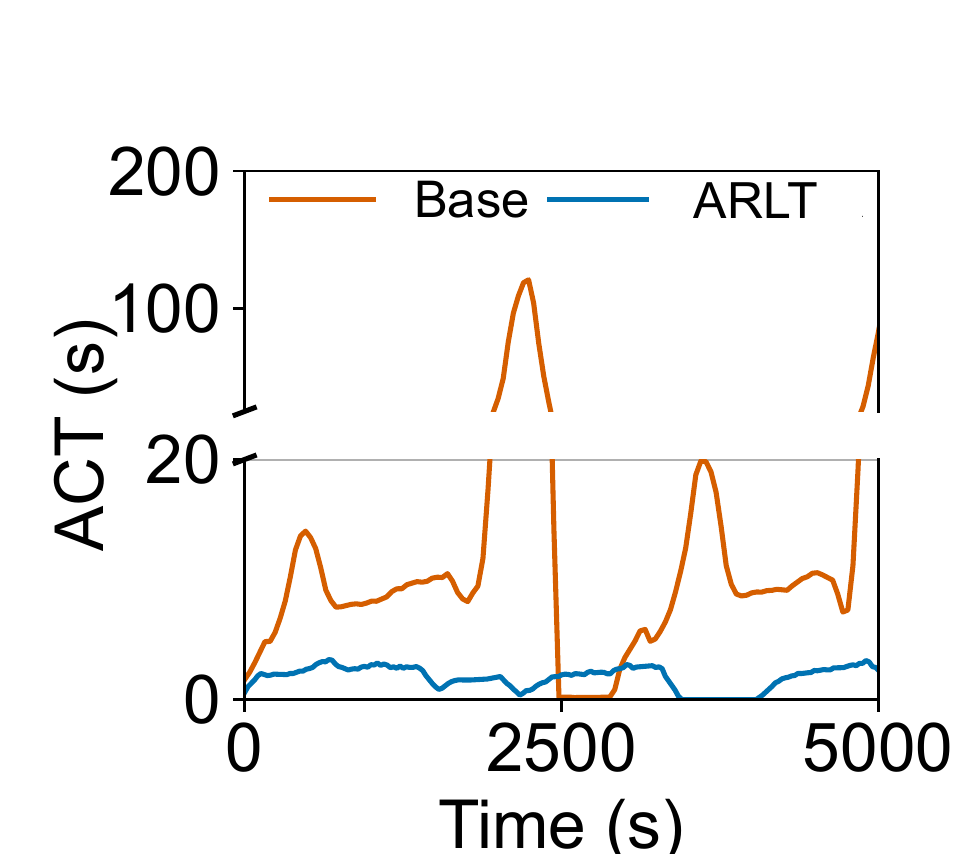}}
        % \vspace{-0.15in}
        \centerline{\small (d) MOPD+DeepSearch.}
        \vspace{-0.1in}
    \end{minipage}
    \caption{ACT under different RL tasks. The numbers and arrows indicate the step durations of RL tasks.}
    \label{fig:eval_specific_lat}
    \vspace{-0.15in}
    \end{figure*}

\subsection{Experiment Setup}
\label{sec:setup}
\paraf{Testbed.} The testbed consists of two components: a GPU cluster running the agentic RL framework, and a set of heterogeneous external resources managed by \sysname.
The RL framework is deployed on a production cluster with up to 48 GPU nodes, each equipped with eight NVIDIA Hopper-architecture GPUs and interconnected via high-bandwidth NVLink and RDMA networks. We adopt VeRL~\citep{sheng2025hybridflow} as the agentic RL framework and equip it with the capacity of external resource invocation, sequence-level and asynchornous \textit{rollout}. 
\sysname is deployed on both CPU and GPU clusters. The CPU cluster contains 15 nodes, each with 256 AMD CPU cores and 2.4 TB of memory. The GPU cluster consists of five nodes, each equipped with eight high-end NVIDIA GPUs and 3 TB of CPU memory. In addition, \sysname manages access to several external API services with enforced quotas, ranging from Google Search to PDF parsing services.

\parabf{Workloads.} We select three agentic RL tasks as workloads in \S\ref{sec:e2e}. \textbf{AI coding} constructs isolated execution environments using CPUs, where environments are invoked intermittently and used to compute reward scores.
We use an in-house dataset following the scaffold of SWEBench~\citep{jimenez2024swebench}, which is designed for production use and is more comprehensive than SWEBench.
\textbf{DeepSearch} continuously accesses external websites to assist LLM-based question answering. Its dataset is based on BrowseComp~\citep{wei2025browsecomp}, and rewards are computed using an LLM-based judge, finetuned GPT-OSS~\citep{agarwal2025gpt}, over generated trajectories. 
GRPO~\citep{shao2024deepseekmath} is adopted for these two workloads.
\textbf{MOPD}~\citep{xiao2026mimo} integrates multiple RL tasks, including agentic tasks, during the \textit{rollout} phase and computes trajectory log-probabilities with respect to corresponding teacher models for alignment.
To demonstrate improvements in mitigating over-provisioning within RL tasks, we additionally run DeepSearch and MOPD concurrently, as both require GPU resources to deploy external services.

\parabf{Baselines.} 
We design workload-specific baselines for the experiments in \S\ref{sec:e2e}, as no existing system uniformly supports all three workloads. For \textbf{AI Coding}, we deploy Kubernetes to manage the CPU cluster. Each trajectory requests the creation of a pod at the beginning of execution, allocating 0.5 CPU per pod to allow limited multiplexing, with an upper bound of four CPUs to prevent congestion. For \textbf{MOPD}, we deploy nine teacher models on the GPU cluster using SGLang~\citep{zheng2024sglang}, allocating four GPUs per model with tensor parallelism (TP). For \textbf{DeepSearch}, the baseline allows for each trajectory to independently perform API calls and retry at most three times when encountering errors or timeout after 600s. The reward service of DeepSearch is also deployed on the GPU cluster with five replicas and degree of TP as 8. As for "MOPD+Search", there are 10 reward services in total and we deploy each model with four GPUs with TP.

\parabf{Metrics.}
To compare the efficiency of \sysname and baselines, we mainly measure the average ACT.
We also report the detailed durations for multiple stages, including LLM generation, tool invocation, and reward computation.
All stage durations are normalized by the total duration of \sysname, so that the normalized durations of \sysname sum up to 1.

\parabf{Other setup.}
In this part, both batch sizes of RL training and external resources capacity are fixed for different workloads: 1280, 2048, 2048, respectively for AI coding, MOPD and DeepSearch. We select Qwen3-32B~\citep{yang2025qwen3} and MiMo-V2-Flash~\citep{xiao2026mimo} for the models of RL training.
Furthermore, the concrete configuration of \sysname encompasses: $depth$ in the scheduling algorithm as 2; scalability and execution durations profiled in advance, only for reward calculation on CPUs and reward model inference on GPUs.

\subsection{End-to-End Performance}
\label{sec:e2e}
% Because GRPO~\citep{} is adopted as the RL algorithm, the concurrency of external invocations is strictly related to current trajectories, i.e. the product of the batch size and the number of samples.

\begin{figure}[t]
    \centerline{\includegraphics[width=0.6\linewidth]{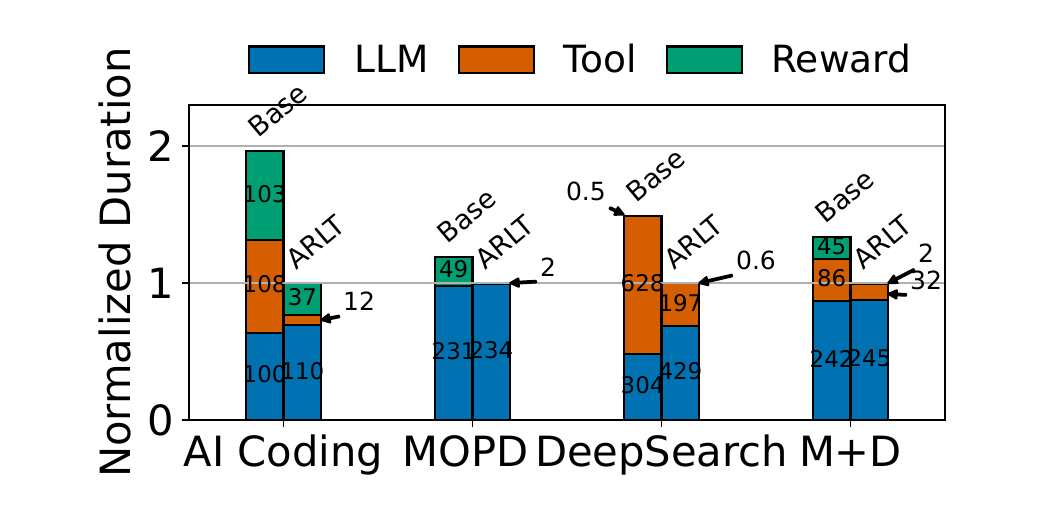}}
    \vspace{-0.3in}
    \caption{End-to-end performance for different RL tasks. Numbers in bars indicate absolute durations (s).}
    \label{fig:eval_e2e}
    \vspace{-0.25in}
    \end{figure}

We evaluate End-to-End performance on multiple workloads to exhibit the generality and efficiency of \sysname. 
Figure~\ref{fig:eval_specific_lat} shows the average ACTs over consecutive small time windows as RL training proceeds under different workloads. We observe that the ACTs under \sysname are consistently lower than those of the baselines. This result indicates that \sysname handles bursty workloads more effectively with the same amount of external resources and reduces ACTs by mitigating over-provisioning and improving the utilization of external resources. Although API calls in DeepSearch are inherently non-scalable, reasonable management of their quotas and concurrency helps avoid errors and retries caused by rate limits or timeouts, thereby reducing ACTs.
% Moreover, due to the bursty nature of invocations in agentic RL training, peak ACTs under \sysname are significantly lower than those of the baselines—by up to xxx×—demonstrating that \sysname handles bursty workloads more effectively with the same amount of external resources by mitigating over-provisioning.

We further report the average duration of ten RL training steps, referred to as \textbf{step duration}, highlighting how \sysname contributes to end-to-end training efficiency improvement. Step durations of both AI Coding and DeepSearch decrease substantially, with improvements of $1.4\times$ and $1.5\times$, respectively.
For AI Coding, the improvement mainly comes from action-level scheduling, which increases the utilization of external CPUs.
For DeepSearch, where API calls are not scalable, the reduction in step duration is mainly attributed to traffic control.
For baseline, frequent API failures cause trajectories to become ineffective, reducing the pass rate of RL training and consequently making the step slow. 
In contrast, due to the specialized pipeline of MOPD, the duration of the \textit{rollout} phase is dominated by the long-tail trajectory rather than the average trajectories.
As a result, the improvement of step duration for MOPD is relatively small. 

Figure~\ref{fig:eval_e2e} provides a detailed breakdown of trajectory durations, further illustrating the improvements brought by \sysname in external invocations.
For AI Coding, the durations of both environment interactions and reward computation decrease significantly, by $9.0\times$ and $2.8\times$, respectively and $4.3\times$ in total.
Specifically, the former benefits from mitigating over-provisioning within trajectories, which reduces the action queuing time, while the latter benefits from elastic allocation of DoP for scalable actions, reducing the execution time.
MOPD benefits from multiplexing teacher models and pooling GPU resources across RL tasks, since trajectories from different RL tasks exhibit diverse and partially interleaved patterns.
In contrast, the baseline can only serve with a fixed number of GPUs, leading to longer execution times.
Notably, in DeepSearch, the reward duration of \sysname is slightly longer than that of the baseline due to restoration overhead, as there is only one model served by the same cluster.
However, in the combined “MOPD+Search” setting, \sysname demonstrates substantial advantages over the baselines, similar to MOPD.

\subsection{System Scalability}
\label{sec:eval_scale}

\begin{figure}[t]
    \begin{minipage}{0.48\linewidth}
    \centerline{\includegraphics[width=\linewidth]{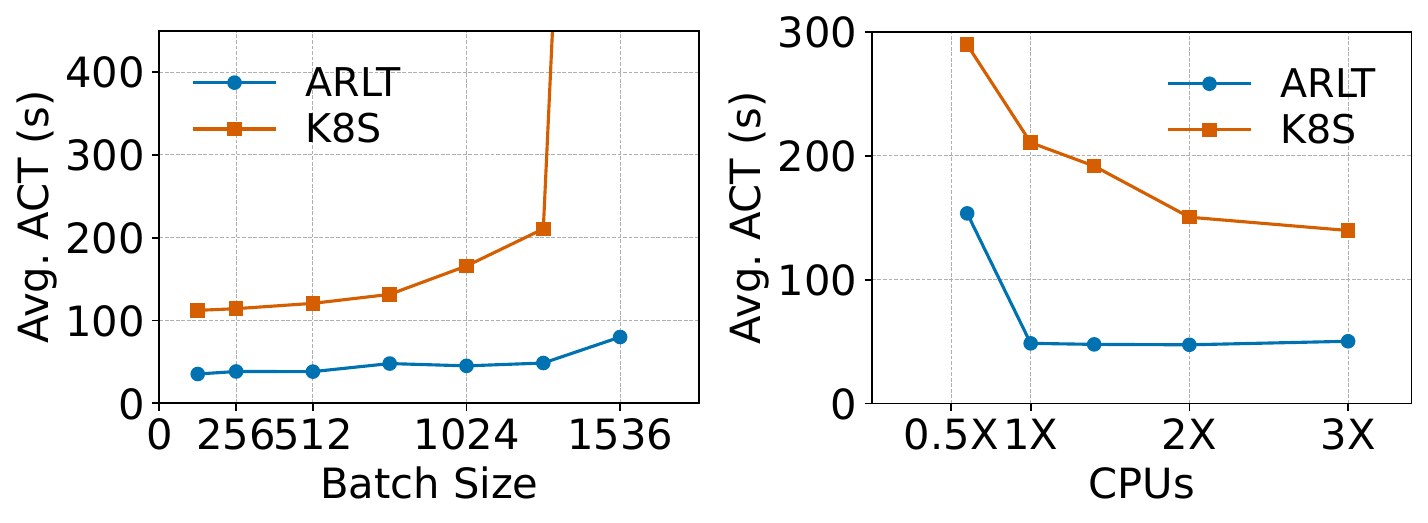}}
    % \vspace{-0.1in}
    \centerline{\small (a) Scalability of CPU.}
    \label{fig:eval_scale_cpu}
    % \vspace{-0.15in}
    \end{minipage}
    \begin{minipage}{0.48\linewidth}
    \centerline{\includegraphics[width=\linewidth]{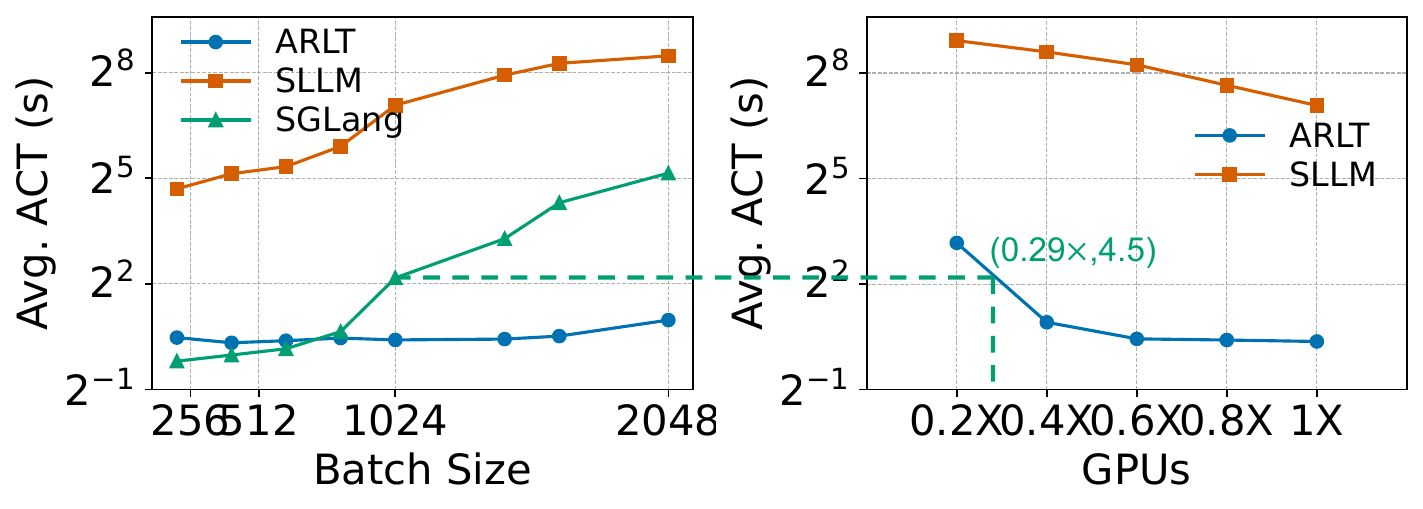}}
    % \vspace{-0.1in}
    \centerline{\small (b) Scalability of GPU.}
    \label{fig:eval_scale_cpu}
    % \vspace{-0.15in}
    
    \end{minipage}
    
    \caption{Scalability of different external resources, in RL batch size and resource capacity.}
    \label{fig:eval_scale}
    \end{figure}

% \begin{figure}[t]
%     \begin{minipage}{0.48\linewidth}
%         \centerline{\includegraphics[width=\linewidth]{figures/Eval_bs_gpu.pdf}}
%         % \vspace{-0.15in}
%         \centerline{\small (a) RL batch size.}
%         % \vspace{-0.1in}
%     \end{minipage}
%     \begin{minipage}{0.48\linewidth}
%         \centerline{\includegraphics[width=\linewidth]{figures/Eval_node_gpu.pdf}}
%         % \vspace{-0.15in}
%         \centerline{\small (b) External resource.}
%         % \vspace{-0.1in}
%     \end{minipage}
%     \caption{Scalability of GPU.}
%     \label{fig:eval_scale_gpu}
%     % \vspace{-0.15in}
%     \end{figure}

We further evaluate the effectiveness of \sysname by scaling both concurrency and resource capacity.
We focus on resources on which actions are scalable, i.e., CPUs and GPUs.
We measure the total ACT per trajectory and report the average across trajectories under different \textbf{batch sizes} and \textbf{resource capacity}. 
Since GRPO is adopted, we treat the number of trajectories in each batch as the RL batch size.

\parabf{Scalability of CPU.}
In Figure~\ref{fig:eval_scale}(a), all experiments are provisioned with 1280 CPU cores across five nodes as external resources. As the RL batch size increases, \sysname reduces the average ACT by $3.1\times$ to $27.7\times$ compared with the baseline.
In particular, when the RL batch size is 1536, the Kubernetes baseline becomes heavily overloaded, triggering frequent queuing timeouts in its control plane due to insufficient resources.
In contrast, \sysname handles this batch size with an acceptable increase in average ACT, i.e., less than $2\times$ compared to the ACT of batch size 128.
The right figure fixes the RL batch size at 1280, which does not fully congest Kubernetes due to the principle of fairness. Under this setting, the baseline remains $1.89\times$ to $4.33\times$ slower than \sysname.
When only 768 CPU cores are available, the improvement becomes smaller, as less resources limit the opportunities for multiplexing across trajectories under action-level scheduling.

% \begin{figure}[t]
    
%     \centerline{\includegraphics[width=\linewidth,trim=0 0 0 40,clip]{figures/gpu_scale.pdf}}
%     \vspace{-0.1in}
%     \caption{Scalability of GPU.}
%     \label{fig:eval_scale_gpu}
%     \vspace{-0.2in}
%     \end{figure}
    
\smallskip \parabf{Scalability of GPU.}
In addition to the SGLang baseline, we include ServerlessLLM~\citep{fu2024serverlessllm} as another baseline, which proposes a Model-as-a-Service (MaaS) abstraction to serve multiple models with a fewer number of GPUs.
Similar to CPUs, experiments in Figure~\ref{fig:eval_scale}(b) Left are conducted on the five-node GPU cluster described in \S\ref{sec:setup}.
Under high RL batch sizes, \sysname significantly outperforms both baselines.
At batch size 2048, \sysname reduces the average ACT by $18.1\times$ compared to SGLang, while ServerlessLLM fails to serve requests at this level of concurrency (where too much timeout failures are unacceptable). 
As a reference, at batch size 1024, \sysname outperforms SGLang and ServerlessLLM by $3.4\times$ and $101.8\times$, respectively.
Notably, SGLang achieves slightly better latency under low concurrency, because of the restoration overhead in \sysname at a low concurrency.

Figure~\ref{fig:eval_scale}(b) Right further highlights \sysname’s advantage in reducing external resource cost under a fixed batch size of 1024. \sysname is able to serve 10 reward services using only 29\% of the GPUs required by the over-provisioned baseline and achieving the same ACT.
Although both \sysname and ServerlessLLM can deploy multiple models with fewer GPUs, ServerlessLLM has substantially higher latency, as it lacks elastic reallocation for reward services and incurs higher system overhead compared to our GPU Manager.

\subsection{Analysis of \sysname}
\label{sec:eval_analyse}
\begin{figure}[t]
        \centerline{\includegraphics[width=0.6\linewidth]{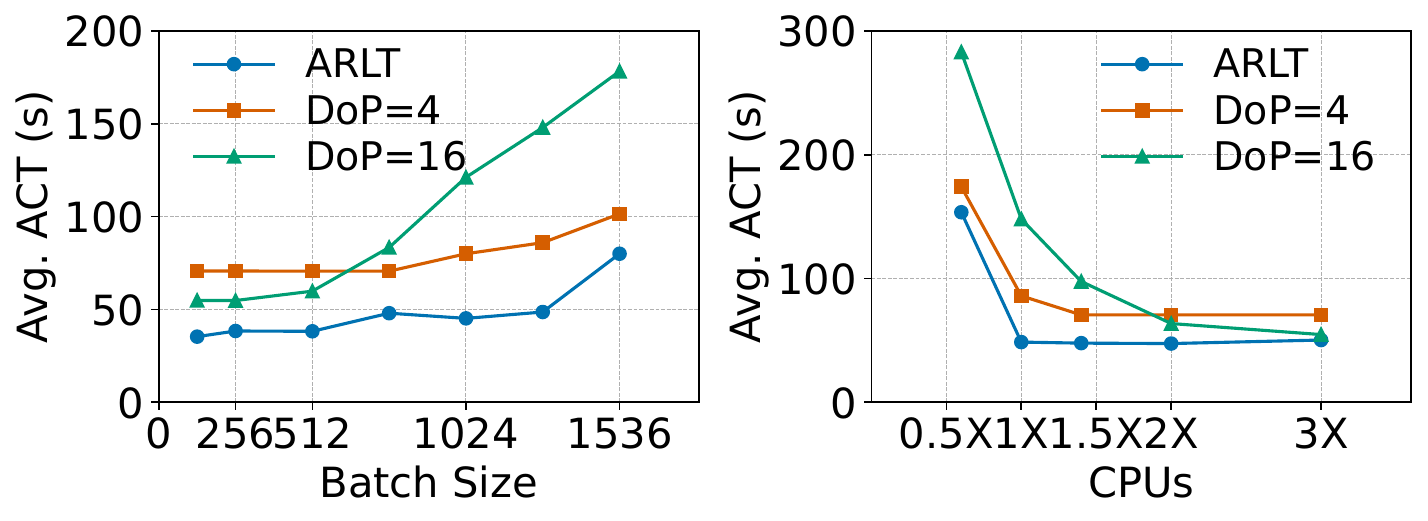}}
        % \vspace{-0.1in}
    \caption{Impact of \sysname scheduling.}
    % \vspace{-0.2in}
    \label{fig:eval_scheduling}
    % \vspace{-0.1in}
    \end{figure}

\paraf{Impact of \sysname Scheduling.}
We conduct an ablation study of the elastic scheduling algorithm on the AI Coding workload. Since the CPU computation workload is highly coupled with LLM outputs which exhibits substantial randomness, we construct a benchmark based on a real RL trace of AI coding and use it to evaluate different scheduling baselines.
In the AI coding task, only reward-calculation actions are CPU-scalable, as they are long-tailed in execution duration and amenable to parallelization.
We replace the elastic scheduling algorithm, that DoP dynamically ranges from 1 to 32, with two fixed DoP baselines (DoP=4 and DoP=16, chosen based on experience). As shown in Figure~\ref{fig:eval_scheduling}, the elastic scheduling algorithm consistently outperforms the fixed-DoP baselines. In particular, it reduces average ACT by $2.0\times$ compared to DoP=4 at batch size 256 and $3.0\times$ compared to DoP=16 at batch size 1280. Similarly, under limited resources (1× CPU cores), it achieves an $1.8\times$ reduction compared to DoP=4. These improvements arise from the algorithm’s ability to elastically allocate resources and adapt strategies according to resource contention.

\parabf{Overhead Analysis.}
We further analyze the system overhead by breaking down ACTs for AI Coding (CPU-intensive) and MOPD (GPU-intensive), as reported in Table~\ref{table:overhead}.
We find that, even if \sysname is heavily congested with batch size 1536 of AI coding, the system overhead still accounts for a small part, less than 3\% compared with the execution duration, while only 2\% for batch size 1280.
% Even under heavy load with batch size 1536 in AI Coding, system overhead accounts for less than 3\% of execution time (2\% at batch size 1280).
For MOPD on GPUs, overhead—primarily due to restoration—represents roughly 25\% of execution time, which cannot be overlooked.
Fortunately, this overhead does not increase explosively with higher concurrency or more severe queueing (with RL batch size 3072), ensuring efficiency and stability under high-concurrency agentic RL training.

\begin{table}[!t] 
\centering
\begin{tabular}{c|cccc} 
\toprule
\makecell[c]{\textbf{Workloads} \\ (bsz)} & 
\makecell[c]{\textbf{Coding} \\ (1280)} &
\makecell[c]{\textbf{Coding} \\ (1536)} & 
\makecell[c]{\textbf{MOPD} \\ (2048)} &
\makecell[c]{\textbf{MOPD} \\ (3072)} \\
\midrule
Exec Dur.                & 0.975 & 1.206 & 0.621 & 0.705 \\
Queue Dur.               & 0.015 & 0.428 & 0.081 & 15.05 \\
Sys. Overhead            & 0.024 & 0.036 & 0.201 & 0.240 \\ 
\bottomrule
\end{tabular} 
% \vspace{0.1in}
\caption{ACTs Breakdown.} 
\label{table:overhead} 
\vspace{-0.2in} 
\end{table}

\iffalse

\begin{figure}[t]
    \centerline{\includegraphics[width=0.5\linewidth]{figures/placeholder.png}}
    \vspace{-0.1in}
    \caption{\sysname overhead.}
    \label{fig:eval_overhead}
    \vspace{-0.15in}
    \end{figure}
    
\parabf{Accuracy of our Modeling}
\zyh{add one evaluation}

\begin{figure}[t]
\centerline{\includegraphics[width=\linewidth]{figures/placeholder.png}}
\vspace{-0.1in}
\caption{Accuracy of \sysname modeling.}
\label{fig:eval_modeling}
\vspace{-0.15in}
\end{figure}
\fi
\section{Related Work}

\paraf{Optimization for RL framework.}
The benefits of LLM RL have drawn great attention to RL framework optimizations~\citep{yao2023deepspeed,sheng2025hybridflow,hu2024openrlhf,zhong2025optimizing,zhong2025streamrl,fu2025areal,he2025history,wang2025distflow,lei2024puzzle,shen2024nemo,han2025asyncflow}.
OpenRLHF~\citep{hu2024openrlhf} and verl~\citep{sheng2025hybridflow} explore the disaggregated and colocated architectures separately.
Some following frameworks propose fusion~\citep{zhong2025optimizing}, pipelining~\citep{zhong2025optimizing}, streaming~\citep{zhong2025streamrl}, and asynchronous execution~\citep{fu2025areal} to improve system efficiency.
Besides, some recent optimizations propose request-level scheduling, balancing, and speculative decoding~\citep{gao2025rollpacker,jinefficient},  to further speed up the rollout phase.
Despite these advancements, existing frameworks often treat external invocations as secondary, either overlooking their resource management or defaulting to static over-provisioning. 
This leads to significant resource waste, particularly because agentic RL tasks involve longer trajectories and more complex environment interactions.

\parabf{Agent resource management.}
The increasing complexity of agentic RL has necessitated more sophisticated management of external resources used for tools and rewards.
Recent efforts like Mars~\citep{zhu2026mars} have addressed reward computation efficiency by deploying it as a standalone service.
Mars leverages a "request-in, batch-out" characteristic to relax request-level latency requirements into batch-level constraints.
While effective for rewards, \sysname provides a more comprehensive solution by establishing a unified action-level platform for all types of tool usage and reward services, including API calls, code execution, and simulators.
% By shifting control from long-lived trajectories to individual atomic interactions (actions), \sysname enables statistical multiplexing across different tasks and reduces resource idle time that a single service may miss.
MegaFlow~\citep{zhang2026megaflow}, SkyRL-Agent~\citep{cao2025skyrl}, and ROCK~\citep{wang2025let} abstract agent service and adopt techniques in cloud computing to manage tools.
By integrating tool characteristics like scalability and sharing pattern, \sysname enables specialized optimization for tool invocations and reward services.

\parabf{Resource auto-scaling.}
Autoscaling is a cornerstone of cloud computing, employing reactive, proactive, or hybrid approaches~\citep{lorido2014review,wu2023xron,wang2024autothrottle,wu2024can,zhang2024jolteon} to maintain service-level objectives (SLOs).
In the LLM domain, systems~\citep{zhang2025blitzscale,yu2025lambda,xiang2025aegaeon,zhao2025seallm} have optimized autoscaling for large-scale inference workloads.
For RL-specific tasks, Mars~\citep{zhu2026mars} introduces history-based scaling and timeout-aware mechanisms to predict and adjust resource requirements at batch boundaries.
\sysname extends these principles by introducing an elastic action-level scheduler designed to handle the bursty and heterogeneous nature of agentic RL actions.
By modeling the scalability of various actions, \sysname minimizes the total ACT and thus, accelerates the RL training process.

\section{Conclusion}
This paper presents \sysname, a unified resource management system for agentic RL that enables fine-grained sharing and elastic orchestration of heterogeneous external resources. By formulating external interactions at the action level, scheduling them elastically and devising heterogeneous resource managers, \sysname enhances overall training efficiency while improving external resource utilization. Extensive evaluation on real-world agentic RL workloads shows that \sysname significantly shortens training step duration, lowers average ACT, and substantially reduces external resource cost, making agentic RL more efficient in cloud environments.

\newpage
\bibliography{main}
\newpage
\appendix
\section{Pseudo Code for ACTs Approximation}
\label{app:pc_approx}
The pseudo code for ACTs approximation is shown as Algorithm~\ref{alg:approx}.

\begin{algorithm}[h!]
\caption{ACTs Approximation.}
\label{alg:approx}
    \begin{footnotesize}
    \begin{algorithmic}[1]
    \Require The waiting queue of actions $W$, the executing queue $E$, the $depth$ of performing approximation of objective
        \Function{getApproximatedObjective}{$C_j, R_j$}
            \State $AC_j \gets W.\text{split}(R_j) - C_j$ \algcomment{Actions to be scheduled the next step}
            \State $exactObj,~newHeap \gets \text{DPArrange}(R_j, C_j)$
            \State $approxObj \gets newHeap.\text{estimate}(AC_j)$
            \State \Return $exactObj + approxObj$
        \EndFunction

        \Statex
        
        \Function{estimate}{$heap, C$}
        \State $obj \gets \inf, ~heap \gets \text{copy}(heap)$
        \For{$d=1 \rightarrow depth$}
        \State $T_0 \gets C[0].\text{getDur}(d),~ts \gets heap.\text{pop}()$
        \State $tmpObj \gets 0,~tmpHeap \gets \text{copy}(heap)$
        \State $tmpObj+=ts+T_0,~tmpHeap.\text{push}(ts+T_0)$
        \For{$i=1 \rightarrow |C|$}
        \State $T_i \gets C[i].\text{getDur}(),~ts \gets heap.\text{pop}()$
        \State $tmpObj+=ts+T_i,~tmpHeap.\text{push}(ts+T_i)$
        \EndFor
        \If{$tmpObj < obj$}~$obj \gets tmpObj$
        \EndIf
        \EndFor
        \State \Return $obj$
        \EndFunction
    \end{algorithmic}
    \end{footnotesize}

\end{algorithm}

\section{Unified DPArrange Algorithm}
\label{app:udp}

Actually, solving the optimal discrete allocation among several scalable candidates and available units is a classical variant of Multi-dimensional Bin Packing problem.
However, due to the heterogeneous topologies of different resource types, the function \texttt{DPArrange} mentioned in \S\ref{sec:design_scheduling} cannot be fully covered by the classical algorithm to solve Multi-dimensional Bin Packing problem. Therefore, we propose a unified DP algorithm that is topology-agnostic, as shown in Algorithm~\ref{alg:unified_dp}, and coordinates with the design of \texttt{DP Operator} of heterogeneous resource managers.

With the topology-agnostic DPArrange algorithm, we are only required to implement several primitives defined in \texttt{Basic DP Operator} for each specialized topology. \texttt{Basic DP Operator} is applicable to the topology of CPU Manager, while we devise a chunk-wise \texttt{Basic DP Operator} for GPU Manager, as shown in Algorithm~\ref{alg:gpu_dp}.

We define a topology-aware DP operator over GPU chunks of sizes \{1,2,4,8\}.
Each DP state is represented by a tuple $(a, b,c,d)$, indicating the number of available chunks of each size, and is linearized into a unique scalar index via a mixed-radix encoding, ensuring a collision-free and finite DP state space.
The start state corresponds to the minimal supported chunk configuration determined by the cluster scale.
Given a task requiring $k$ GPUs, the predecessor operator greedily decomposes $k$ using available chunks from larger to smaller sizes, allowing chunk splitting while preserving power-of-two constraints.
If the decomposition is feasible, the operator transitions to the encoded residual chunk state; otherwise, the transition is invalid.
This operator enables dynamic programming to reason explicitly about GPU topology while remaining compatible with standard one-dimensional DP formulations.

\begin{algorithm}[t!]

\caption{Topology-Agnostic DPArrange}
\label{alg:unified_dp}
\begin{footnotesize}
\begin{algorithmic}[1]
\Require 
    A set of scalable tasks $C = \{c_1, \dots, c_m\}$,
    available resource units $U$,
    supported unit sets $\mathcal{S} = \{S_1, \dots, S_m\}$,
    resource manager $\mathcal{M}$, and its DP operator $\mathcal{O}$
\Ensure
    Optimal execution time $T^\star$,
    allocation $\{k_1, \dots, k_m\}$

\State $n \gets \min\{\textsc{End}(U),~\textsc{End}(\mathcal{S})\}$
\If{$n < m$}
    \State \Return \textbf{error} \algcomment{Each task requires at least one unit}
\EndIf

\State Initialize DP table $dp[i][j] \gets \infty$ for $0 \le i \le m,~0 \le j \le n$
\State Initialize choice table $choice[i][j] \gets 0$
\State $dp[0][0] \gets 0$

\State $start_{prev} \gets 0$
\State $start_{cur} \gets \mathcal{O}.\textsc{Start}(S_1)$

\For{$i = 1 \rightarrow m$}
    \For{$j = start_{cur} \rightarrow n$}
        \ForAll{$k \in S_i$} \algcomment{Units assigned to task $c_i$}
            \State $j' \gets mathcal{O}.\textsc{Prev}(j, k)$
            \If{$j' < start_{prev}$ \textbf{or} $\neg \mathcal{O}.\textsc{IsValid}(j', S_{1:i-1})$}
                \State \textbf{continue}
            \EndIf
            \State $T_i(k) \gets \mathcal{O}.\textsc{GetDuration}(c_i, k)$
            \If{$dp[i-1][j'] + T_i(k) < dp[i][j]$}
                \State $dp[i][j] \gets dp[i-1][j'] + T_i(k)$
                \State $choice[i][j] \gets k$
            \EndIf
        \EndFor
    \EndFor
    \State $start_{prev} \gets start_{cur}$
    \State $start_{cur} \gets mathcal{O}.\textsc{Start}(S_{1:i+1})$
\EndFor

\State $j \gets n$
\For{$i = m \rightarrow 1$}
    \State $k_i \gets choice[i][j]$
    \State $j \gets mathcal{O}.\textsc{Prev}(j, k_i)$
\EndFor

\State \Return $dp[m][n],~\{k_1, \dots, k_m\}$

\Statex
\Statex \textbf{Basic DP Operator}

\Function{Start}{$\mathcal{S}$}
    \State \Return $\sum_i \min(S_i)$
\EndFunction

\Function{End}{$X$}
    \If{$X$ is a unit list}
        \State \Return $|X|$
    \Else
        \State \Return $\sum_i \max(S_i)$
    \EndIf
\EndFunction

\Function{Prev}{$j, k$}
    \State \Return $j - k$
\EndFunction

\Function{IsValid}{$r, X$}
    \If{$r < 0$}
        \State \Return \textbf{false}
    \EndIf
    \If{$X = \emptyset$}
        \State \Return $(r = 0)$
    \EndIf
    \ForAll{$u \in X_1$}
        \If{$u \le r$ \textbf{and} $\textsc{IsValid}(r - u, X_{2:})$}
            \State \Return \textbf{true}
        \EndIf
    \EndFor
    \State \Return \textbf{false}
\EndFunction

\end{algorithmic}
\end{footnotesize}
\end{algorithm}

\begin{algorithm}[t!]
\caption{GPU-Topology-Aware DP Operator}

\label{alg:gpu_dp}
\label{alg:gpu_dp_operator}
\begin{footnotesize}
\begin{algorithmic}[1]

\Require
Maximum available chunk counts $(N_1, N_2, N_4, N_8)$ for chunk sizes $\{1,2,4,8\}$

\vspace{0.3em}
\Statex \textbf{State Representation:}
\Statex A state is $(a,b,c,d)$ where
\Statex $a \in [0,N_1],~b \in [0,N_2],~c \in [0,N_4],~d \in [0,N_8]$

\vspace{0.5em}
\Function{Encode}{$a,b,c,d$}
    \State \Return $a
    + (N_1+1)\cdot b
    + (N_1+1)(N_2+1)\cdot c
    + (N_1+1)(N_2+1)(N_4+1)\cdot d$
\EndFunction

\vspace{0.3em}
\Function{Decode}{$j$}
    \State $a \gets j \bmod (N_1+1)$
    \State $j \gets \lfloor j / (N_1+1) \rfloor$
    \State $b \gets j \bmod (N_2+1)$
    \State $j \gets \lfloor j / (N_2+1) \rfloor$
    \State $c \gets j \bmod (N_4+1)$
    \State $j \gets \lfloor j / (N_4+1) \rfloor$
    \State $d \gets j$
    \State \Return $(a,b,c,d)$
\EndFunction

\vspace{0.5em}
\Function{Start}{$\mathcal{S}$}
    \Statex $\mathcal{S}$: supported unit sets of all tasks
    \State $(a_{\min},b_{\min},c_{\min},d_{\min}) \gets$ minimal required chunks inferred from $\mathcal{S}$
    \State \Return \textsc{Encode}$(a_{\min},b_{\min},c_{\min},d_{\min})$
\EndFunction

\vspace{0.5em}
\Function{IsValid}{$j$}
    \State $(a,b,c,d) \gets \textsc{Decode}(j)$
    \If{$a<0$ \textbf{or} $b<0$ \textbf{or} $c<0$ \textbf{or} $d<0$}
        \State \Return \textbf{false}
    \EndIf
    \If{$a>N_1$ \textbf{or} $b>N_2$ \textbf{or} $c>N_4$ \textbf{or} $d>N_8$}
        \State \Return \textbf{false}
    \EndIf
    \State \Return \textbf{true}
\EndFunction

\vspace{0.5em}
\Function{Prev}{$j, k$}
    \State $(a,b,c,d) \gets \textsc{Decode}(j)$
    \State $need \gets k$

    \State $use_d \gets \min\bigl(d, \lfloor need/8 \rfloor\bigr)$
    \State $need \gets need - 8 \cdot use_d$

    \State $use_c \gets \min\bigl(c, \lfloor need/4 \rfloor\bigr)$
    \State $need \gets need - 4 \cdot use_c$

    \State $use_b \gets \min\bigl(b, \lfloor need/2 \rfloor\bigr)$
    \State $need \gets need - 2 \cdot use_b$

    \State $use_a \gets \min(a, need)$
    \State $need \gets need - use_a$

    \If{$need > 0$}
        \State \Return $\bot$ \algcomment{Not enough chunks to satisfy $k$}
    \EndIf

    \State $a' \gets a - use_a$
    \State $b' \gets b - use_b$
    \State $c' \gets c - use_c$
    \State $d' \gets d - use_d$

    \State \Return \textsc{Encode}$(a',b',c',d')$
\EndFunction

\end{algorithmic}
\end{footnotesize}
\end{algorithm}

\end{document}